\documentclass[twocolumn,trackchanges]{aastex63}

\usepackage{amsmath}
\usepackage{multirow}
\usepackage{enumitem}
\usepackage{graphicx}
\usepackage{color,hyperref}

\begin{document}

\title{Mapping the Galactic Disk with the LAMOST and Gaia Red Clump Sample \uppercase\expandafter{\romannumeral7}: the Stellar Disk Structure Revealed by the Mono-abundance Populations}

\author{Zheng Yu}
\affiliation{College of Physics, Hebei Normal University, Shijiazhuang 050024, China;
  liji@hebtu.edu.cn~}

\author{Ji Li}
\affil{College of Physics, Hebei Normal University, Shijiazhuang 050024, China;
  liji@hebtu.edu.cn~}

\author{Bingqiu Chen}
\affiliation{South-Western Institute for Astronomy Research, Yunnan University, Kunming 650500, China}

\author{Yang Huang}
\affil{South-Western Institute for Astronomy Research, Yunnan University, Kunming 650500, China}

\author{Shuhua jia}
\affiliation{College of Physics, Hebei Normal University, Shijiazhuang 050024, China;
  liji@hebtu.edu.cn~}

\author{Maosheng Xiang}
\affiliation{Max-Planck Institute for Astronomy, Konigstuhl, D-69117, Heidelberg, Germany}

\author{Haibo Yuan}
\affil{Department of Astronomy, Beijing Normal University, Beijing 100875, China}

\author{Jianrong Shi}
\affil{Key Laboratory of Optical Astronomy, National Astronomical Observatories, Chinese Academy of Science, Beijing 100012, China}
\affil{School of Astronomy and Space Science, University of Chinese Academy of Sciences, Beijing 100049, China}
\author{Chun Wang}
\affil{Department of Astronomy, Peking University, Beijing
100871, China.}

\author{Xiaowei Liu}
\affil{South-Western Institute for Astronomy Research, Yunnan University, Kunming 650500, China}

\begin{abstract}
  Using a sample of 96,201 primary red clump (RC) stars selected from the LAMOST and Gaia surveys, we investigate the stellar structure of the Galactic disk. The sample stars show two separated sequences of high-[$\alpha$/Fe] and low-[$\alpha$/Fe] in the [$\alpha$/Fe]-[Fe/H] plane. We divide the sample stars into five mono-abundance populations (MAPs) with different ranges  of [$\alpha$/Fe] and [Fe/H], named as the high-[$\alpha$/Fe], high-[$\alpha$/Fe] \& high-[Fe/H], low-[Fe/H], solar, high-[Fe/H] MAPs respectively. We present the stellar number density distributions in the $R-Z$ plane, and the scale heights and scale lengths of the individual MAPs by fitting their vertical and radial density profiles. The vertical profiles, the variation trend of scale height with the Galactocentric radius, indicate that there is a clear disk flare in the outer disk both for the low-[$\alpha$/Fe] and the high-[$\alpha$/Fe] MAPs. While the radial surface density profiles show a peak radius of 7\,kpc and 8\,kpc for the high-[$\alpha$/Fe] and low-[$\alpha$/Fe] MAPs, respectively. We also investigate the correlation between the mean rotation velocity and metallicity of the individual MAPs, and find that the mean rotation velocities are well separated and show different trends between the high-[$\alpha$/Fe] and the low-[$\alpha$/Fe] MAPs. At last, we discuss the character of the high-[$\alpha$/Fe] \& high-[Fe/H] MAP and find that it is more similar to the high-[$\alpha$/Fe] MAP either in the radial and vertical density profiles or in the rotation velocity.
\end{abstract}


\section{Introduction}
  \label{sec:intro}
  The Galactic stellar disk is the most prominent part of our own Milky Way and contains about three quarters of all Galactic stars \citep{Rix  Bovy(2013)}, the special distributions as well as the chemical abundances and kinematic properties of stars in the Galactic disk provide important information for probing the spatial structure and understanding the formation and evolution of the Galactic disk. So that it is very important to investigate the  correlations between spatial structure and abundances as well as the kinematical structures of the Galactic disk stars \citep{van der Kruit  Freeman(2011)}.

  Since the thick disk was first discovered by \citet{Gilmore  Reid(1983)}, who found that the vertical stellar distribution was well fitted by two exponentials with different scale heights. Using star counts in the solar neighbourhood, the geometrical dichotomy structure had been observed by several subsequent studies (e.g., \citealt{Robin et al.(1996)}; \citealt{Juric et al.(2008)}; \citealt{Kordopatis et al.(2011)}; \citealt{Cheng et al.(2012)}). These studies confirmed that the Galactic disk consists of two stellar populations, i.e. the distinct thin and thick components. Generally, the thick disk is thought to have a larger scale height (h$_Z$) than the thin disk. For example, \citet{Juric et al.(2008)} suggested a thin disk with h$_Z$ $=$ 300\,pc and a thick disk with h$_Z$ $=$ 900\,pc based on SDSS (the Sloan Digital Sky Survey, \citealt{York et al.(2000)}) photometric survey analysis, these values agree well with the first results presented by \citet{Gilmore  Reid(1983)}. \citet{Kordopatis et al.(2011)} derived an h$_Z$ = 216 $\pm$ 13\,pc and an h$_Z$ = 694 $\pm$ 45\,pc for the thin and thick disks, respectively. While there are some inconsistencies in the values of scaled length (h$_R$) presented in literature. \citet{Juric et al.(2008)} found that the thin disk has a smaller scale length (h$_R$ = 2.9\,kpc) compared to the thick disk (h$_R$ = 3.6\,kpc), while \citet{Cheng et al.(2012)} showed the opposite trend (see also \citealt{Bensby et al.(2011)}), i.e. the scale length of h$_R$ = 3.4\,kpc for the thin disk is larger than that of h$_R$ = 1.8\,kpc for the thick disk using a sample of 5,620 main-sequence turnoff stars from the SEGUE \citep{Yanny et al.(2009)} survey. Recently, \citet{Wang et al.(2018)} found the radial density profile of the disk shows three separated segments with different scale lengths of 2.12 $\pm$ 0.26, 1.18 $\pm$ 0.08, and 2.72\,kpc at $R$ $<$ 11, 11 $\leq$ $R$ $\leq$ 14, and $R$ $>$ 14\,kpc, respectively. Although there are variations in the estimated sizes of the thin- and thick- disk scale heights and lengths, the spatial dichotomy structure of the thin- and thick-disk populations have been established in or outside the solar neighbourhood.

  On the other hand, the two distinct stellar populations have also been identified in chemical abundance space. Over the last twenty years, a great of high-resolution spectroscopic studies have observed a bimodal distribution of stars in the [$\alpha$/Fe] (here $\alpha$ refers to the average of Mg, Si, Ca, and Ti abundances ) versus [Fe/H] plane (e.g., \citealt{Fuhrmann(1998), Fuhrmann(2008)}; \citealt{Prochaska et al.(2000)}; \citealt{Bensby et al.(2003), Bensby et al.(2004), Bensby et al.(2014)}; \citealt{Reddy et al.(2006)}; \citealt{Adibekyan et al.(2011, Adibekyan et al.(2012), Adibekyan et al.(2013)}; \citealt{Haywood et al.(2013)}). These works show that the thick-disk stellar populations are in general more metal-poor and $\alpha$-enhanced, older ages and hotter kinematics than most stars of the thin-disk populations, a particularly striking result is the clear distinction of [$\alpha$/Fe] between the thin- and thick-disk stars, which either have been identified kinematically (e.g., \citealt{Bensby et al.(2014)}) and then shown to be chemically distinct, or identified chemically and then shown to present different kinematical properties (e.g., \citealt{Adibekyan et al.(2013)}). However, most of the high-resolusion observations were confined to the solar neighborhood with dozens of or hundreds of sample stars selected kinematically (e.g., \citealt{Bensby et al.(2014)}), the small size in star number and narrow range in spatial coverage prevent us to understand the complete structure of the whole disk of the Galaxy. Meanwhile, kinematical selections are significantly biased and can introduce bias or mixing between the selected thin and thick disk samples \citep{Hinkel et al.(2014)}. \citet{Navarro et al.(2011)} argued that it is better to identify stars with different populations based on their elemental abundances rather than other properties such as kinematics. \citet{Adibekyan et al.(2013)} applied a purely chemical analysis approach based on the [$\alpha$/Fe] vs. [Fe/H] plot of about 850 FGK solar neighborhood dwarfs observed with the HARPS high-resolution spectrograph to separate Galactic stellar populations into the thin disk, thick disk, and high-$\alpha$ metal-rich. They found that the gradient of mean rotation velocity with metallicity is also different for the chemically different stellar populations, i.e. the thin disk shows a negative gradient while the thick disk shows a positive gradient, but in turn the kinematic properties of the thin- and thick-disk stars are also mixed in the Toomre diagram. Recently, using the dimensionality reduction technique t-SNE (t-distributed stochastic neighbour embedding, \citet{Anders et al.(2018)} reanalyzed a sample of 1,111 solar-vicinity FGK stars observed with the HARPS spectrograph (R $\sim$ 115,000) to further dissect stellar chemical abundance space in the solar neighbourhood. They found the high- and low-[$\alpha$/Fe] stellar sequences, as well as the high-[$\alpha$/Fe] metal-rich population. To better understand the complex abundance structure of the Galactic stellar disk, it is crucial to map the distribution of elements throughout the disk, beyond the solar neighborhood.

  In recent years, observations from the low- and medium-resolution large-scale spectroscopic surveys RAVE \citep{Steinmetz et al.(2006)}, LAMOST/LEGUE \citep{Zhao et al.(2012), Deng et al.(2012)}, and SDSS/SEGUE \citep{Yanny et al.(2009)}, together with the high-resolution surveys Gaia-ESO \citep{Gilmore et al.(2012)}, APOGEE \citep{Majewski et al.(2017)}, HERMES/GALAH \citep{De Silva et al.(2015), Martell et al.(2017)}, and K2-HERMES \citep{Wittenmyer et al.(2018)} confirmed the two chemically distinct stellar populations with low- and high-[$\alpha$/Fe] ratios across the whole Galactic disk (e.g., \citealt{Lee et al.(2011), Anders et al.(2014), Nidever et al.(2014), Mikolaitis et al.(2014), Hayden et al.(2015), Bovy et al.(2016), Lian et al.(2020)}). While \citet{Bovy et al.(2012a), Bovy et al.(2012b)} argued that there is no clear separation between structurally thin and thick disks, but rather a smooth transition when they investigated the scale-height of the individual mono-abundance populations from the SEGUE DR7 \citep{Abazajian et al.(2009)}. \citet[hereafter B16]{Bovy et al.(2016)} suggested an useful methodology to look at the complex correlations between the spatial structure and stellar population of the Galactic disk, by separately determining disk structure of so-called mono-abundance populations (MAPs). Using a sample of 14,699 red-clump stars (RCs) from the APOGEE-RC catalog, B16 investigated the distributions of RCs in [$\alpha$/Fe] vs. [Fe/H] plane. Their results confirm the structural dichotomy both in chemical and spatial distributions, with a larger scale-height and shorter scale-length for the high-[$\alpha$/Fe] MAPs compared with the low-[$\alpha$/Fe] MAPs. Particularly, they found that the low-[$\alpha$/Fe] MAPs show clear evidence of flaring (the disk thickness increases with radius), while the high-[$\alpha$/Fe] MAPs does not display any flaring.

  Considering the analyses of data from spectroscopic surveys have led to conflicting results about the vertical characteristics of the Milky Way disk (e.g., \citealt{Bovy et al.(2012a), Bovy et al.(2012b), Bovy et al.(2016)}), stellar ages appear to be a better discriminator for identifying stellar populations. Based on the photometric data obtained from the K2 Galactic Archaeology Program (K2 GAP, \citealt{Stello et al.(2015), Stello et al.(2017)}), \citet{Rendle et al.(2019)} analysed the ages of red giant stars from two campaign fields C3 and C6 of K2 GAP and found a clear bimodality in the age distributions for both the Kepler and the K2 fields, with distinct young and old age at 5 and 14\,Gyr for a sample of red giants with age uncertainties  $<$ 35 per cent (in their Figure 13). In addition, they found clear associations of this age dichotomy with the populations defined geometrically and/or chemically to each peak: 5\,Gyr$-$low-$\alpha$, $\mid Z\mid$ $\leq$ 1.0\,kpc (thin disc); 14\,Gyr$-$high-$\alpha$, $\mid Z\mid$ $>$ 1.0\,kpc (thick disc). Using a large sample of stars with very precise metallicity measurements from spectroscopic surveys like the Galactic Archaeology for HERMES (GALAH, \citealt{De Silva et al.(2015)}), K2-HERMES (a GALAH-like survey dedicated to K2 follow-up, \citealt{Wittenmyer et al.(2018)}), and APOGEE \citep{Majewski et al.(2017)} surveys, \citet{Sharma et al.(2019)} estimated a mean age of about 10\,Gyr for the thick disk stars, and found that most of the old thick disk stars are restricted to 5\,kpc $<$ $R$ $<$ 7\,kpc and 1\,kpc $<$ $\mid Z\mid$ $<$ 2\,kpc, which is in agreement with the traditional characters of an old $\alpha$-enhanced thick disk.

  The Large Sky Area Multi-Object Fiber Spectroscopic Telescope (LAMOST) is a reflecting Schmidt telescope \citep{Cui et al.(2012)}, which can take 4,000 spectra in a single exposure at the low resolution of R = 1,800 or the medium resolution of R = 7,500 \citep{Zhao et al.(2012)}. The LEGUE (LAMOST Experiment for Galactic Understanding and Exploration) spectroscopic survey is a major component of the LAMOST project, it was designed to study the structure of Galactic halo and disk components \citep{Deng et al.(2012), Luo et al.(2015)}. From the pilot survey started in October 2011 up to the end of 2020, including the finished Phase 1 Survey and the ongoing Phase 2 Survey, the LAMOST survey have released seven times dataset as the DR1 to DR7. The DR1 $\sim$ DR6 have been released to the public, while the DR7 is only released to domestic astronomers and international collaborators at present. The DR7 v1.2 dataset was acquired from October 2011 to June 2019 with a total of 14.48 million spectra, including 10.6 million low-resolution spectra, 1.01 million medium-resolution non-time domain spectra, and 2.87 million medium-resolution time domain spectra (these information come from the LAMOST official website http://dr7.lamost.org/v1.2). It also includes a catalogues of about 6.93 million stars' spectral parameters derived from high quality spectra with S/N $>$ 10. The large amount of stellar spectra will allow us to systematically investigate the spatial density, Galactocentric rotation velocity and velocity ellipsoid, and chemical abundance of stars as a function of position in the Galaxy.

  In this paper, we use a sample of primary red clump stars (RCs) presented by \citet[hereafter as Paper \uppercase\expandafter{\romannumeral1}]{Huang et al.(2020)} to explore the three dimensional structure of the Galactic stellar disk utilizing the mono-abundances methodology of B16. This RC sample was selected from the LAMOST DR4 and Gaia DR2 \citep{Gaia Collaboration et al.(2018), Lindegren et al.(2018)}, which includes about 140,000 RCs with the accurate measurements of distance, proper motions and stellar atmospheric parameters (effective temperature $T_{\rm{eff}}$, surface gravity log $g$ and metallicity [Fe/H]), and $\alpha$-element to iron abundance ratio, etc. Several works have published based on this RC sample, for example, \citet{Li et al.(2020)} studied the kinematic signature of the Galactic warp; \citet{Sun et al.(2020)} investigated the origin of the ``young'' [$\alpha$/Fe]-enhanced stars of the Galactic disk; \citet{Wang et al.(2020)} analyzed the amplitude evolution of the stellar warp and found that it is induced by the nongravitational interaction over the disk models. These works unravel the stellar structure and assemblage history of the Galactic disk in different aspects.

  The outline of this paper is as follows. In Section \ref{sec:data}, we introduce the data of our sample stars. In Section \ref{sec:constraction}, we constructed a number density map of stars in the Galactic disk. We calculate the scale heights and scale lengths for the individual MAPs in Section \ref{sec:result}. In Section \ref{sec:discussion}, we give some discussions about our results. Finally, we present our conclusions in Section \ref{sec:conclusion}.

\section{Data}\label{sec:data}
  As standard candles widely distributed across the entire Galactic disk, RCs are excellent tracers to explore the structure of the Galactic disk. Our sample stars are selected from the LAMOST-RC sample presented by Paper \uppercase\expandafter{\romannumeral1}. The sample covers a large volume of the Galactic disk with 4 $\leq$ $R$ $\leq$ 20\,kpc, $|Z|$ $\leq$ 5\,kpc, and $-$20$^{\circ}$ $\leq$ $\varphi$ $\leq$ 50$^{\circ}$, where $R$ is the Galactocentric distance in the cylindrical coordinate system, $Z$ is the vertical distance from the Galactic plane, and $\varphi$ is the Galactocentric azimuth in the direction of Galactic rotation. The stellar parameters of the RCs, including the effective temperature ($T_{\rm{eff}}$), surface gravity(log $g$), metallicity ([Fe/H]), line-of-sight velocity ($V_{r}$), and elemental abundance [$\alpha$/Fe], are determined with the pipeline of LAMOST Stellar Parameter at Peking University (LSP3; \citealt{Xiang et al.(2015), Xiang et al.(2017a)}; \citealt{Li et al.(2016)}). The second release of value-added catalogues for the LAMOST Spectroscopic Survey of the Galactic Anticentre (LSS-GAC DR2, \citealt{Xiang et al.(2017b)}) presents the typical uncertainties for $T_{\rm{eff}}$, log $g$, [Fe/H], $V_{r}$, and [$\alpha$/Fe] are 100\,K, 0.25\,dex, 0.10\,dex, 5\,km/s, and 0.05\,dex, respectively. The LAMOST-RC sample also provided the accurate distances, masses, and ages of the RCs, with the typical uncertainties of 5$-$10 percent, 15 percent, and 30 percent, respectively (see details in Paper \uppercase\expandafter{\romannumeral1}).

  To ensure the reliability and authenticity of our results on the Galactic disk structure revealed from the spatial distributions of our RCs sample, it is necessary to do a correction of the selection effect for the whole RCs sample of Paper \uppercase\expandafter{\romannumeral1}. Here we adopt the selection function $S$ presented by \citet{Chen et al.(2018)} to correct the selection effects. Based on the source of origin, $S$ mainly consists of two parts. The first part, quantified by $S_{1}$, characterizes the LAMOST target selection strategy. The second part, quantified by $S_{2}$, characterizes the selection effects due to the observational quality, data reduction, and parameter determination. Each star in our sample is weighted by $1/S$ in the color-magnitude diagram (CMD). The LSS-GAC targets are selected from the photometric catalogues of the Xuyi Schmidt Telescope Photometric Survey of the Galactic Anticentre (XSTPS-GAC; \citealt{Zhang et al.(2013)}; \citealt{Liu et al.(2014)}) and the AAVSO Photometric All-Sky Survey (APASS; \citealt{Henden et al.(2016)}). For more details about the calculations of the selection function, one is suggested to see the literature of \citet{Chen et al.(2018)}. Considering the individual spectrographs of LAMOST have an average area of 1.2\,deg$^{2}$ \citep{Chen et al.(2018)}, which is slightly larger than the size of 1\,deg$^{2}$ for the boxes used for target selection. Consequently, some stars were  discarded from the observed sample in the operation of correction to sample selection. Finally, we obtained a sample of 96,201 primary RCs from the LAMOST-RC catalog.

\begin{figure}
\includegraphics[height=0.3\textheight,width=0.45\textwidth,clip=]{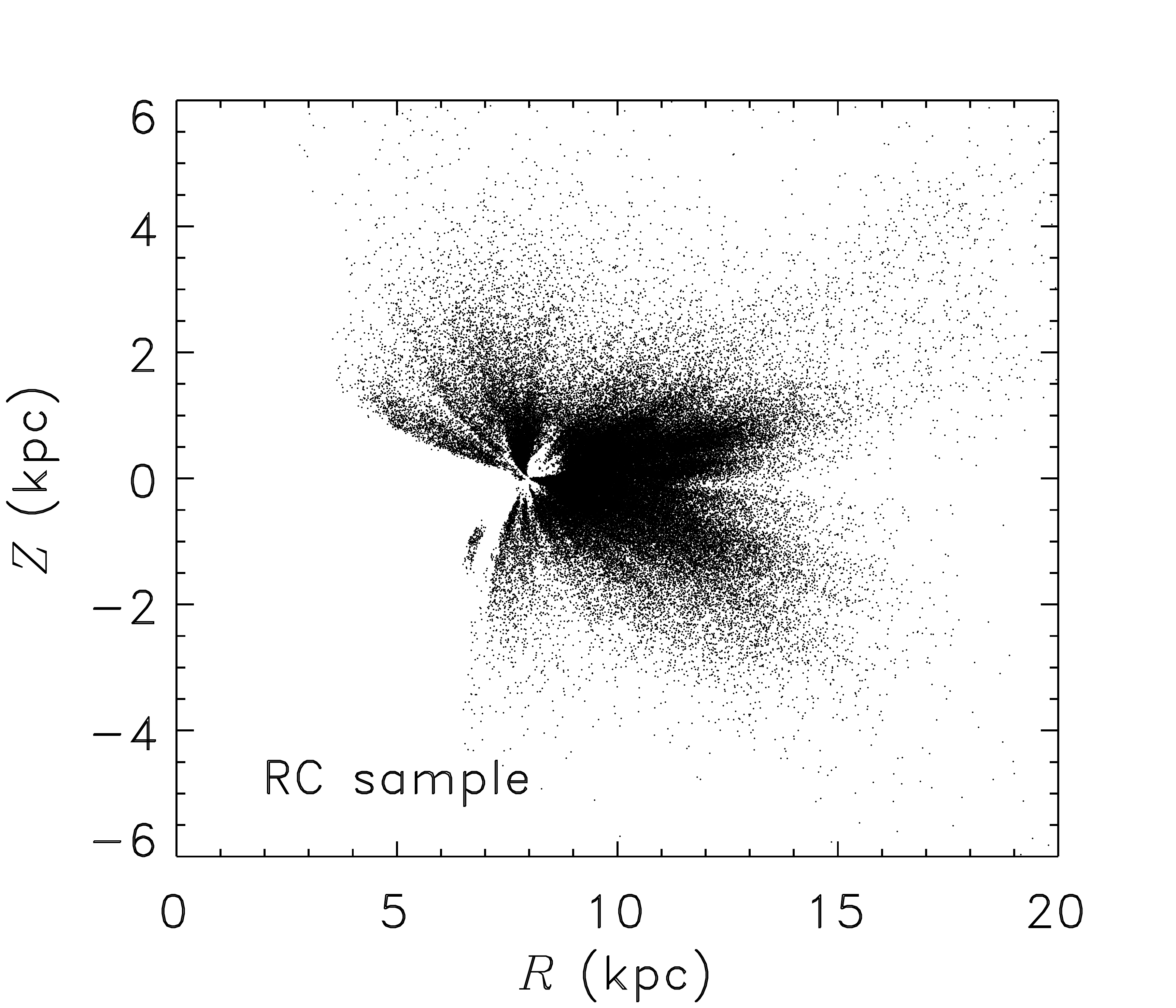}
\caption{The distribution of the RC sample in $R - Z$ plane. The Sun is located at ( $R$ , $Z$ ) = ( 8, 0.025 )\,kpc.}
\label{fig:f1}
\end{figure}

  \begin{figure}
  \includegraphics[height=0.3\textheight,width=0.5\textwidth,clip=]{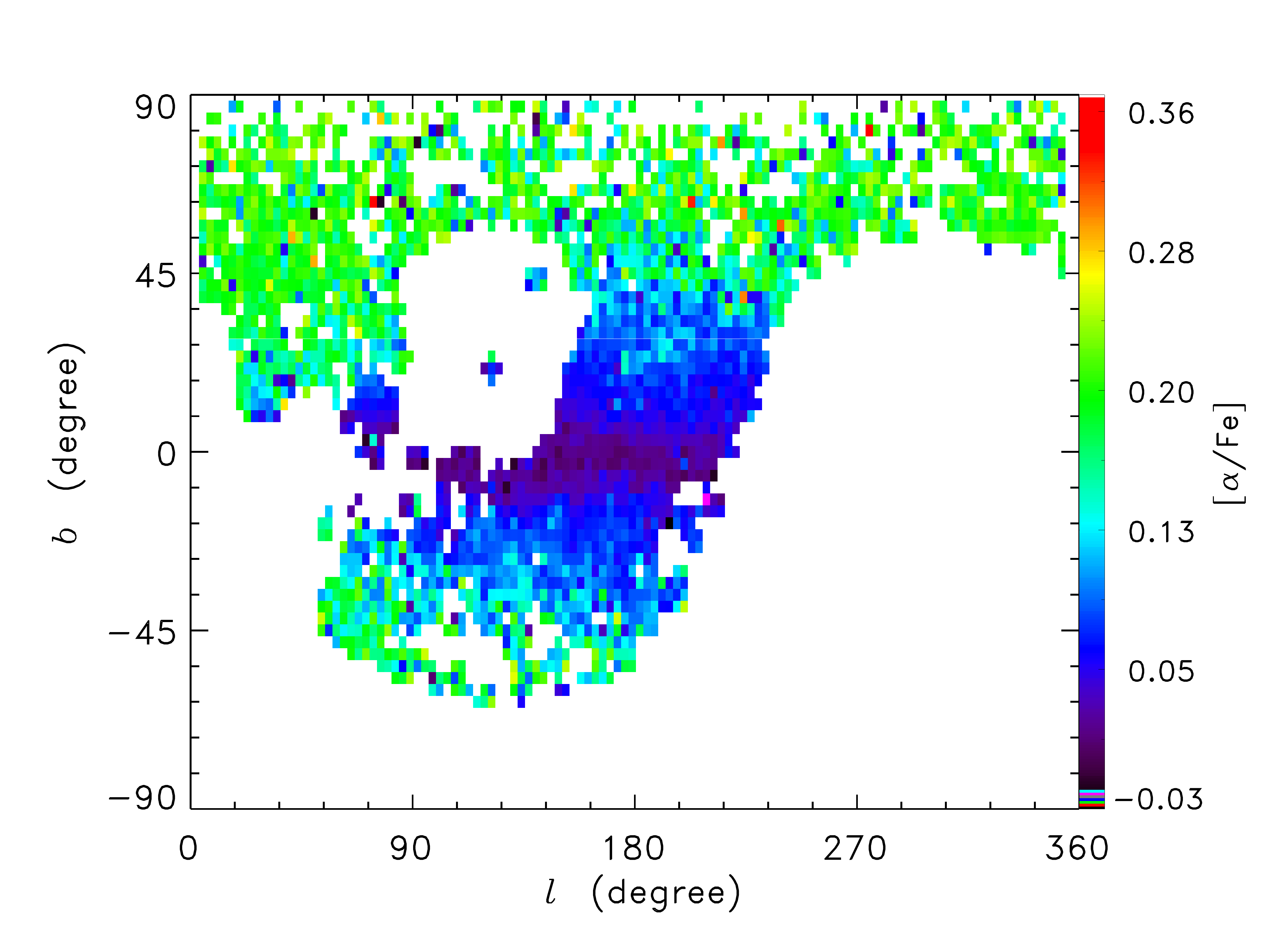}
 \caption{The footprints of our RCs in a Galactic coordinate system. The data are divided into patches of 3$^\circ  \times$ 3$^\circ$ .
  \label{fig:f2}}
 \end{figure}

\begin{figure}
  \includegraphics[height=0.35\textheight,width=0.5\textwidth,clip=]{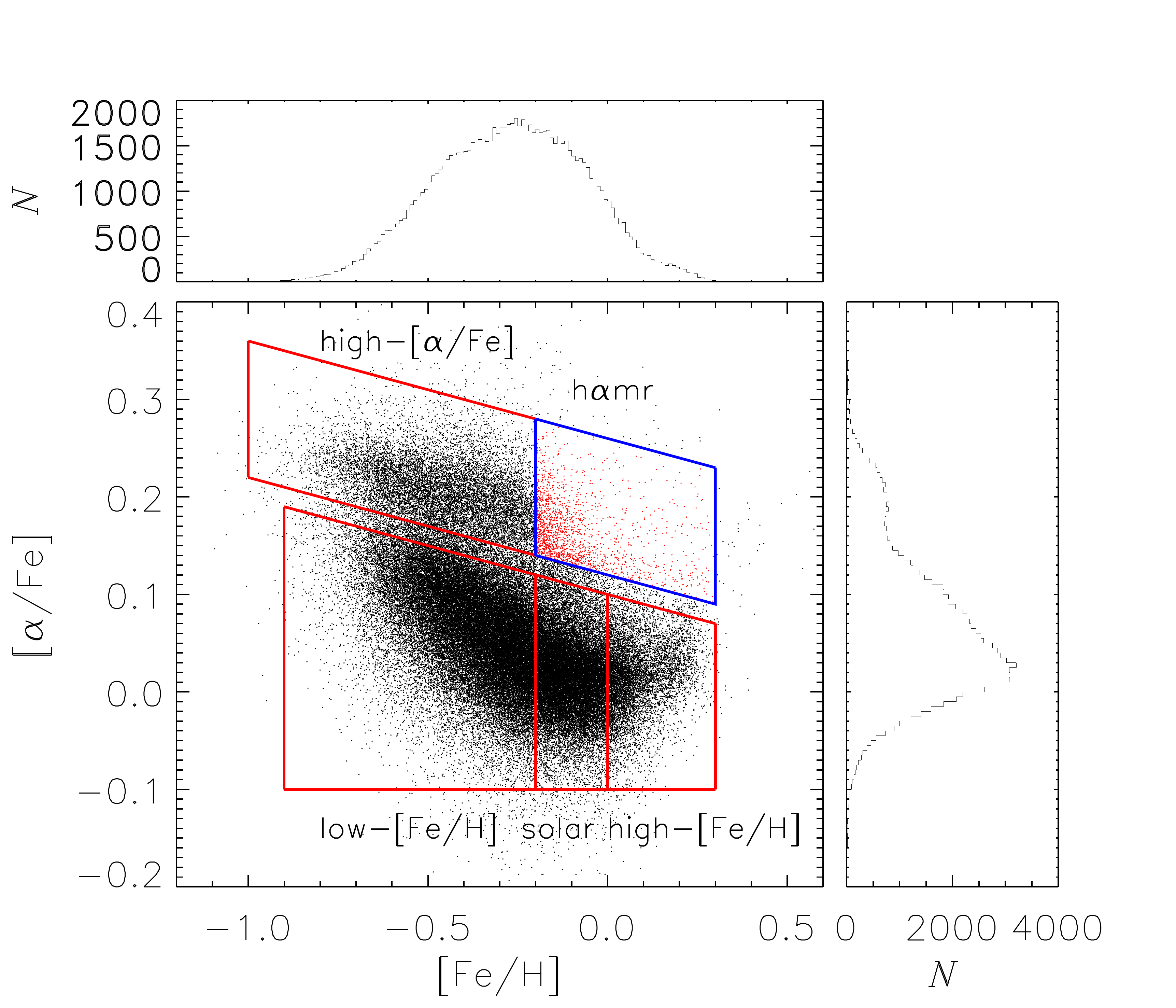}
  \caption{Distribution of the RC sample stars in the plane of [$\alpha$/Fe]-[ Fe/H]. The red lines delineate the boundaries of the four MAPs that we study in Section \ref{sec:data}, which we denote with the given moniker. The blue lines delineate the boundaries of the h$\alpha$mr MAP. \label{fig:f3}}
  \end{figure}

  Figure \ref{fig:f1} displays the spatial distribution of our RC stars in the $R-Z$ plane, we adopt the position of the Sun is at $R$ = 8\,kpc and $Z$ = 25\,pc \citep{Juric et al.(2008)} in our work. It shows that our RCs are distributed over a large space with 4 $\leq$ $R$ $\leq$ 20\,kpc, $|Z|$ $\leq$ 5\,kpc, but only a small number of stars located at the very outer disk ( $R$ $\geq$ 20\,kpc ). Figure \ref{fig:f2} presents the footprints of our RCs in a Galactic coordinate system, we note that this distribution is consistent with that of the whole sample stars of LAMOST DR4 (see Figure 2 in \citealt{Xiang et al.(2017b)}), which means that our RC sample has similar spatial completeness to the whole sample of LAMOST DR4. Figure \ref{fig:f3} shows the distribution of RCs in [$\alpha$/Fe]-[Fe/H] plane. It exhibits a bimodality in the plane, showing two distinct high-$\alpha$ and low-$\alpha$ sequences as presented by the APOGEE RC stars of B16, in their Figure 5. Therefore, we can use the same methodology as used by B16 for their APOGEE RC stars to investigate the disk structure based on spatial distributions of the two distinct stellar populations of our RCs.

  In order to compare the results with B16, we divide artificially our whole RCs into four broad mono-abundances populations named as the high-[$\alpha$/Fe], low-[Fe/H], solar, and high-[Fe/H] MAPs, respectively, as displayed in Figure \ref{fig:f3}, and the later three MAPs also collectively referred to as low-[$\alpha$/Fe] MAPs in the following. To reduce the mixing between the high- and low-[$\alpha$/Fe] sequences as far as possible, we separate them with a narrow isolation belt with a width of 0.03\,dex in [$\alpha$/Fe] and oblique along the [Fe/H] with a slop of $-$0.1. The contour of the high-[$\alpha$/Fe] MAP is a rhomboid with [Fe/H] from $-$1.0 to 0.$-$2\,dex. But the low-[Fe/H], solar, and high-[Fe/H] MAPs are continuous in metallicity, corresponding to the metallicity ranges of $-$0.9 $\sim$ $-$0.2\,dex, $-$0.2 $\sim$ 0.0\,dex, and 0.0 $\sim$ +0.3\,dex, respectively. The number of stars in the four MAPs of high-[$\alpha$/Fe], low-[Fe/H], solar, and high-[Fe/H] are 11,137, 45,403, 25,484, and 8,137, respectively. In the meantime, we also selected an intriguing MAP of high-[$\alpha$/Fe] \& high-[Fe/H] (h$\alpha$mr) that is surrounded by the blue lines in Figure \ref{fig:f3}, which we will discuss in subsection \ref{sec:five map}.

\section{Construction of the number density maps}\label{sec:constraction}
  To map the distributions of the stellar number density, we divide the sample stars into different bins of the distance modulus $\mu$ for each MAP, and assign the size of $\mu$ bin is 0.5 for the high-[Fe/H] MAP and 0.3 for the rest three MAPs.

  The stellar number density $H(d_{i})$ in a volume element $\Delta$$V(d_{i})$ with a distance bin d$_{i}$ is defined as

  \begin{equation}
  H(d_{i}) = N(d_i)/ \Delta V(d_i),
  \label{8}
  \end{equation}

  where $N(d_{i})$ is the star count in the volume bin $\Delta$$V(d_{i})$ at distance d$_{i}$, which has been corrected with our selection function. $\Delta$$V(d_{i})$ is the volume given by
  \begin{equation}
  \Delta V(d_i) =\frac{\omega}{3}(\frac{\pi}{180})^2(d^3_{i2}-d^3_{i1}),
  \label{9}
  \end{equation}
  where $\omega$ denotes the area of the field (unit in deg$^2$ ), $d_{i1}$ and $d_{i2}$ are the lower distance limit ($d_{i}$ $-$ bin/2) and upper distance limit ($d_{i}$ + bin/2) of each $d_{i}$bin, respectively .

\begin{figure*}
\begin{center}
\includegraphics[height=0.22\textheight,width=0.4\textwidth,clip=]{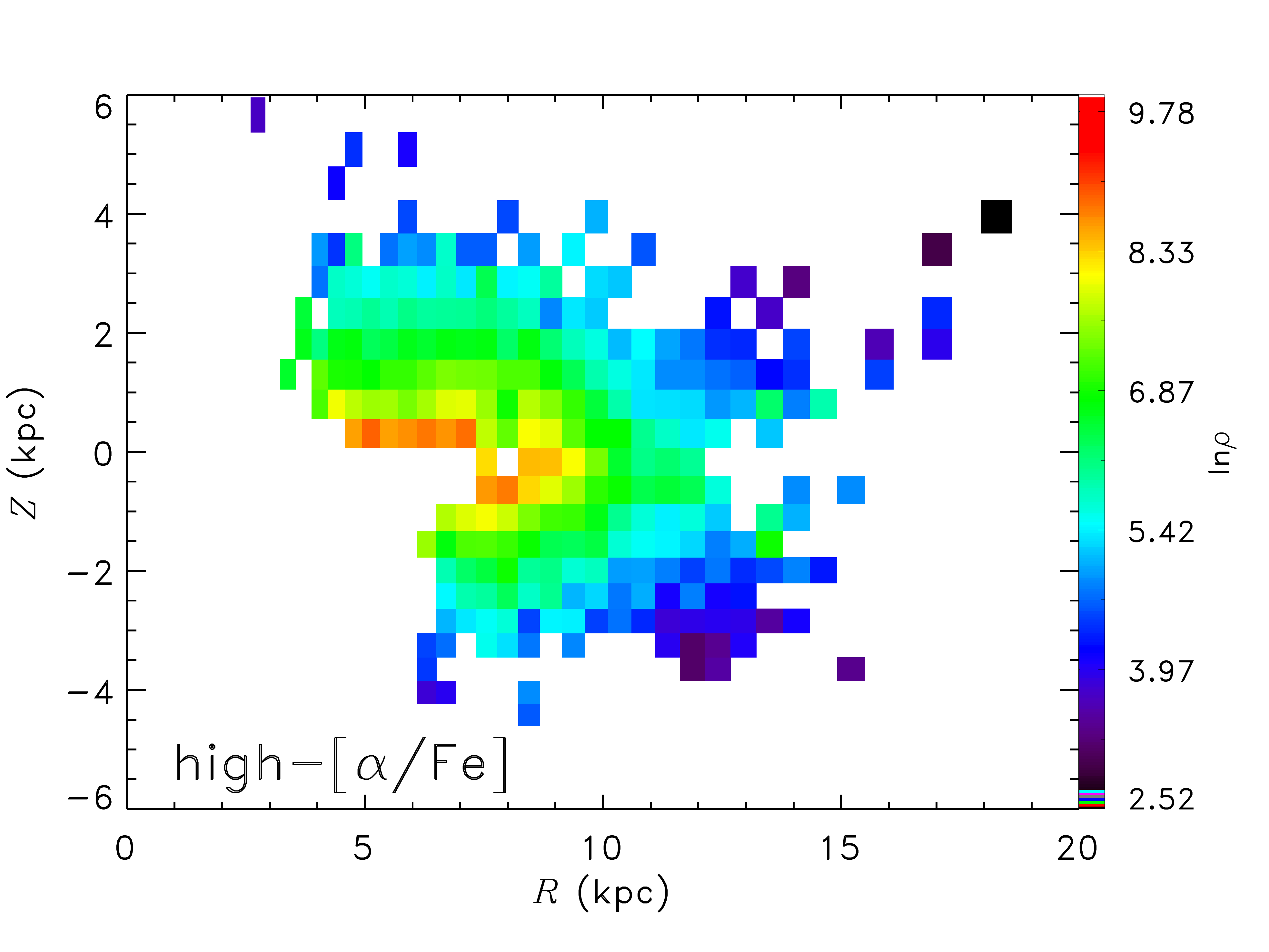}
\includegraphics[height=0.22\textheight,width=0.4\textwidth,clip=]{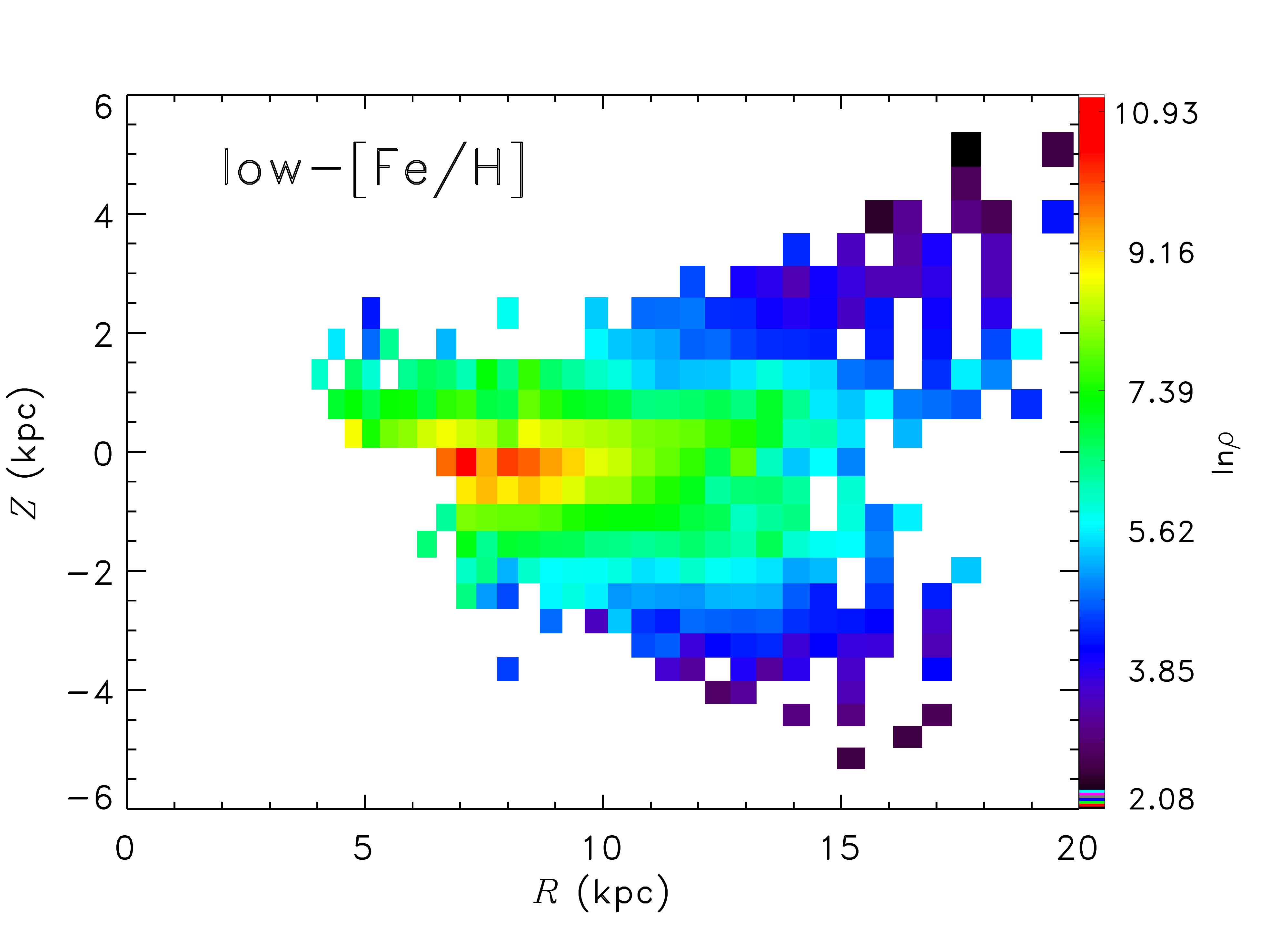}\\
\includegraphics[height=0.22\textheight,width=0.4\textwidth,clip=]{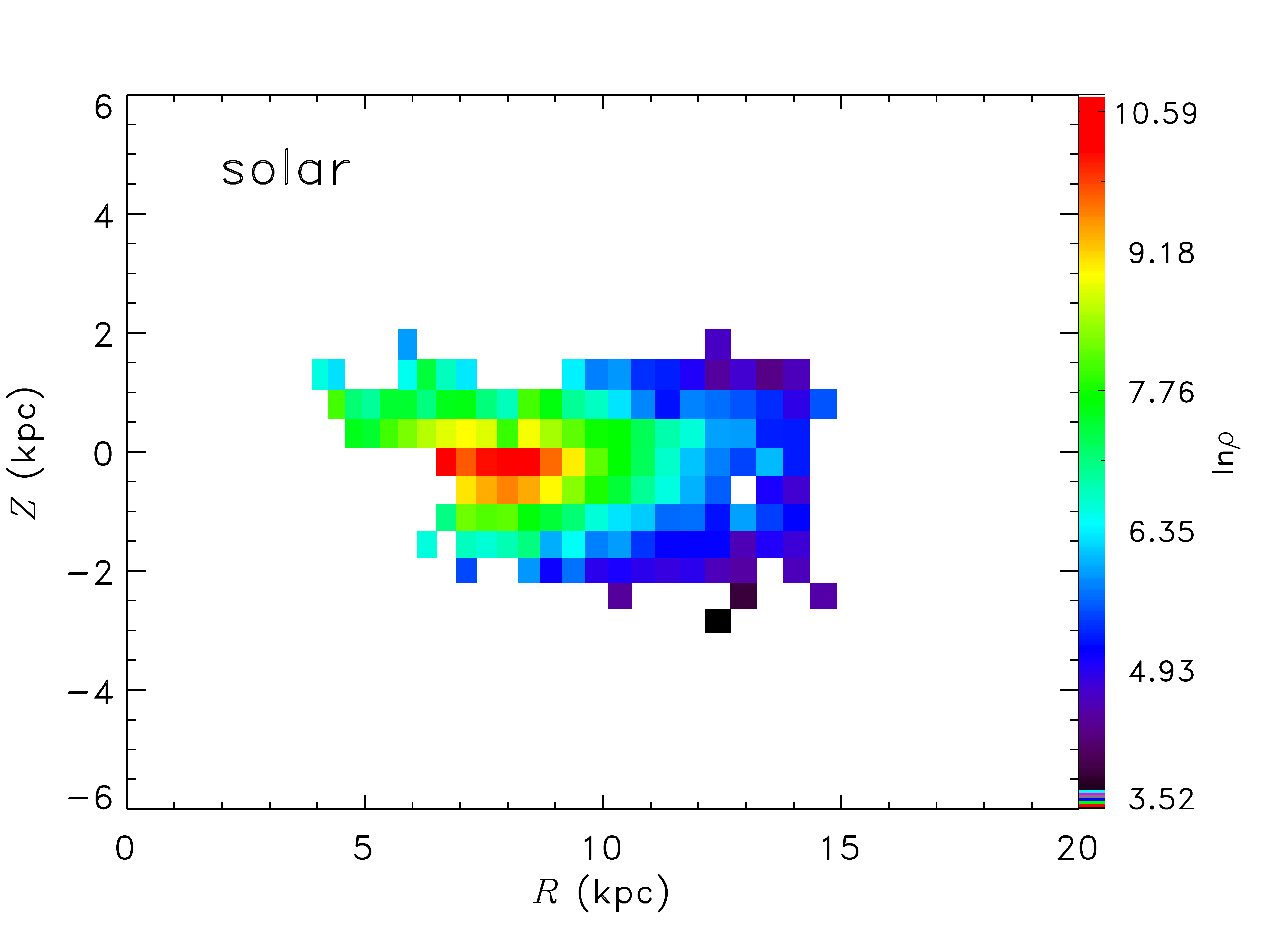}
\includegraphics[height=0.22\textheight,width=0.4\textwidth,clip=]{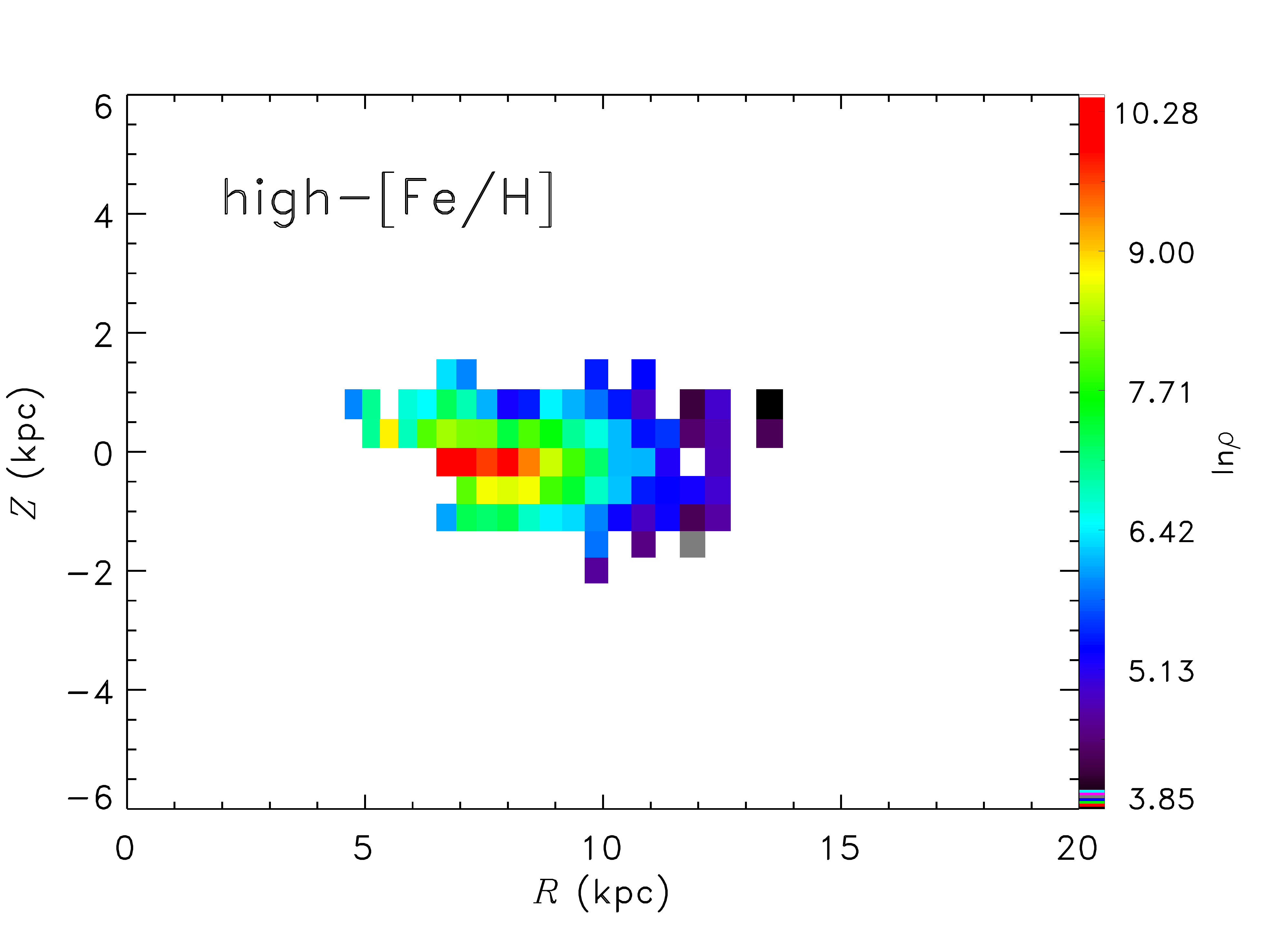}
\caption{Stellar number density distributions in the $R-Z$ plane deduced from four broad subsamples after corrected for selection biases. The colour encodes the logarithmic mean stellar number density(ln$\rho$). \label{fig:f4}}
\end{center}
\end{figure*}

  By the above calculations, the stellar number densities for the individual volume bins at different distance bins were determined. Then we divided them into narrow bins in $R-Z$ plane to map the distributions of stellar number density for the individual MAPs as displayed in Figure \ref{fig:f4}, the right color bar ln$\rho$ represents the logarithm of the stellar number density and we have taken account of the corrections of selection effect to calculate the number density of the individual $\mu$ bins as described in Section \ref{sec:constraction}. As displayed in Figure \ref{fig:f4}, we note that the high-[$\alpha$/Fe] and low-[Fe/H] MAPs are throughout all the space ranges of the whole RC sample distributions with the 4\,kpc $<$ $R$ $<$\,20 kpc and $-$5\,kpc $<$ $Z$ $<$ 5\,kpc. But stars of the solar and the high-[Fe/H] MAPs are mainly located in a relatively small space with 5\,kpc $<$ $R$ $<$\,15 kpc and $-$2\,kpc $<$ $Z$ $<$ 2\,kpc.

\section{Result}\label{sec:result}

\subsection{The vertical profile}\label{sec:hz}
  we assume the vertical profile of the stellar number density $H$ in each MAP of the RCs could be modeled as an exponential decrease with the increase of the distance from the mid-plane ($Z$), which is given by,
  \begin{equation}
  H = H_{0}\rm{exp}(-Z/h_Z),
  \label{3}
  \end{equation}
  where $H_{0}$ is the local stellar number density in the mid-plane of the Galactic disk, h$_{Z}$ is the scale height of one sub-sample MAP. Since the spatial distribution of the stars on the Galactic plane is approximately symmetrical, we take the absolute value of the $Z$ for stars located in the south sky area as the $Z$ value used in the calculation of Equation \ref{3}.

  To explore the vertical stellar density profiles of the individual MAPs, we divided each of the four subsamples into small radial bins $\vartriangle$$R_{i}$, and assume each $\vartriangle$$R_{i}$ bin has a radially constant scale height $h_{Z} (R_{i})$, $R_{i}$ is the corresponding projected Galactocentric radius. The widths of the individual $\vartriangle$$R_{i}$ bins for each MAP are different according to the distributions of stellar number densities. The dividing criteria is to ensure the bin size as small as possible to obtain a better resolution in the radial direction, in the meanwhile, we need to keep as many as possible stars in each bin so that we can have good coverage in the vertical direction. Those bins without enough stars (less than 5 density points) have been ignored in the whole analysis.

  \begin{figure*}
  \begin{center}
  \includegraphics[height=0.4\textheight,width=0.23\textwidth,clip=]{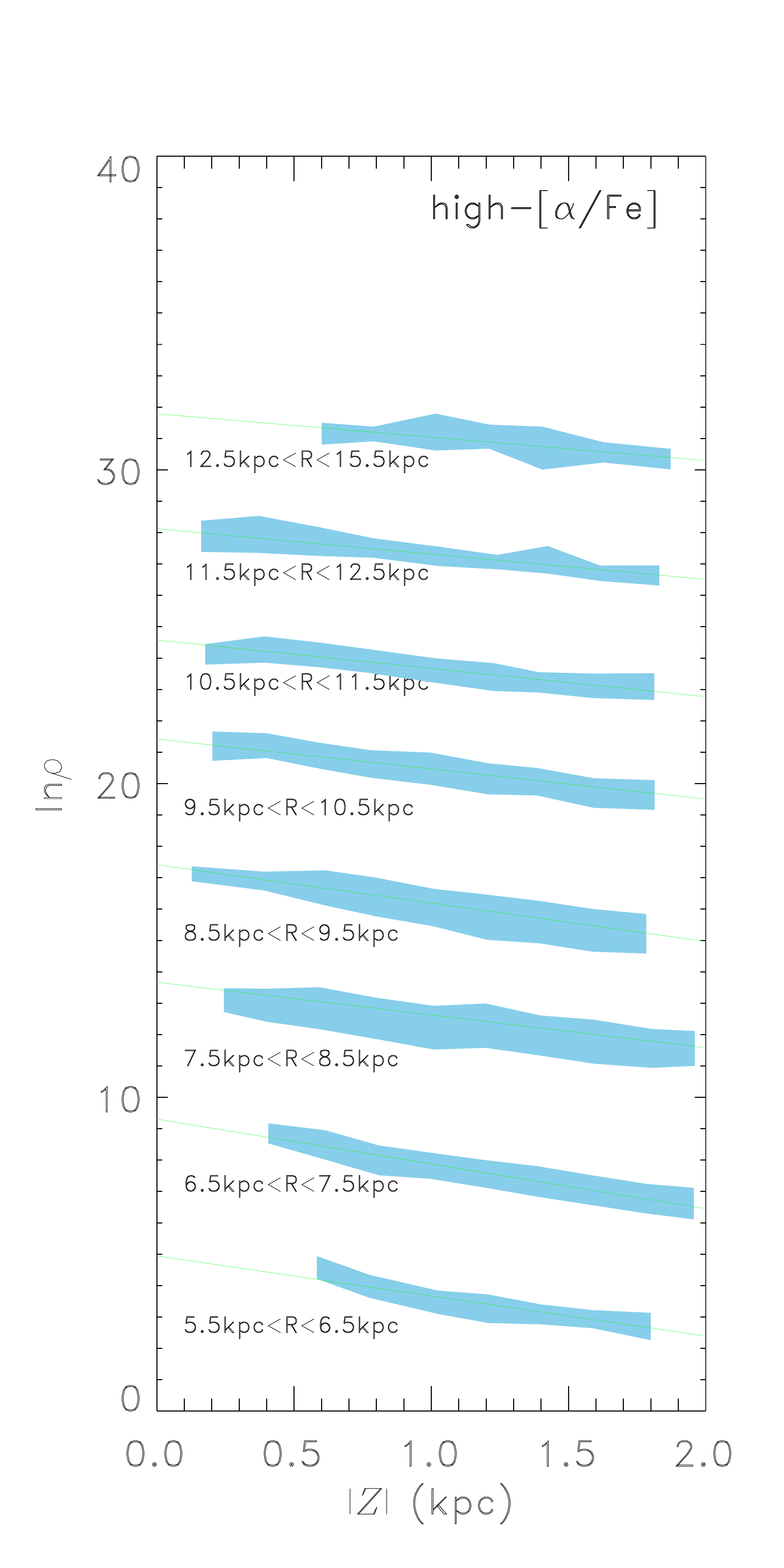}
  \includegraphics[height=0.4\textheight,width=0.23\textwidth,clip=]{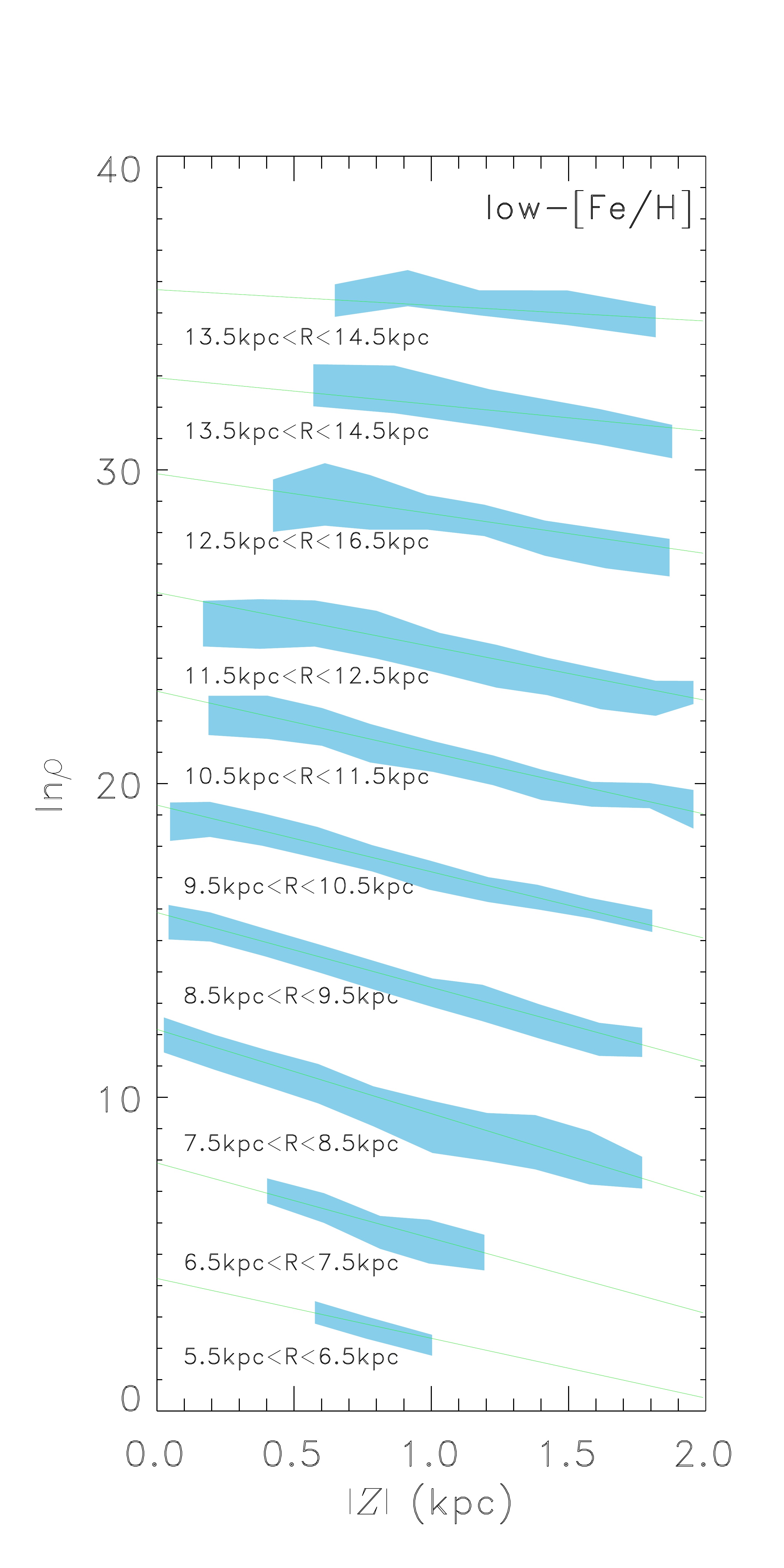}
  \includegraphics[height=0.4\textheight,width=0.23\textwidth,clip=]{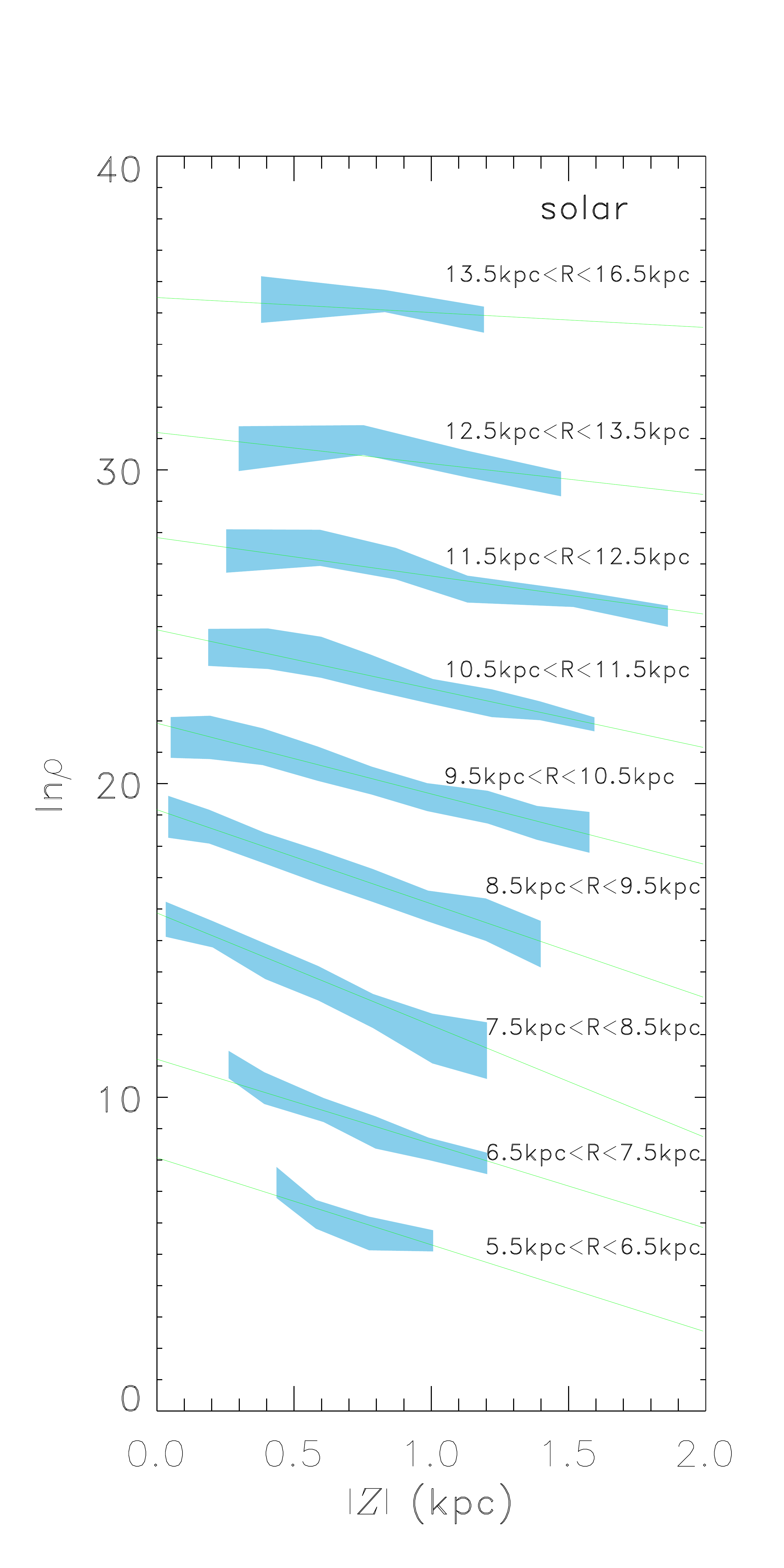}
  \includegraphics[height=0.4\textheight,width=0.23\textwidth,clip=]{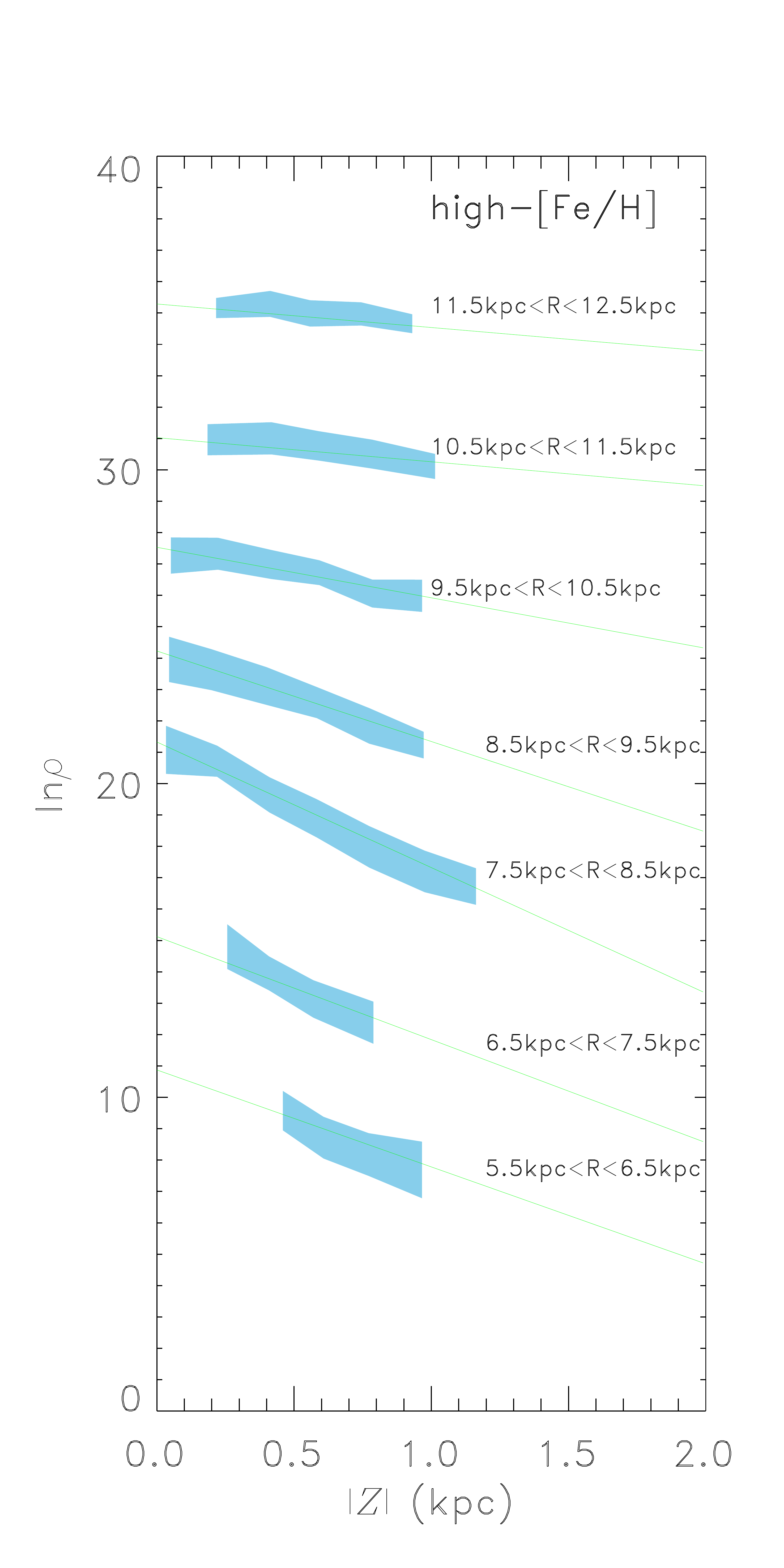}
  \caption{The distributions of the stellar number densities in different radial radius bins along with the absolute value of $Z$ and fitting results with a single exponential for the individual MAPs, as labeled in the panels. The vertical axis ln$\rho$ is the stellar number density in natural logarithm, and the horizontal axis $|Z|$ absolute value of the distance away from the Galactic plane. The shadow regions indicate the distribution range of star number density, and the green lines are the best fit using the least square method. Some arbitrary offsets in the vertical direction of each panel have been applied to separate the different bins of the radial radius. \label{fig:f5}}
  \end{center}
  \end{figure*}
\begin{figure}
  \includegraphics[height=0.3\textheight,width=0.45\textwidth,clip=]{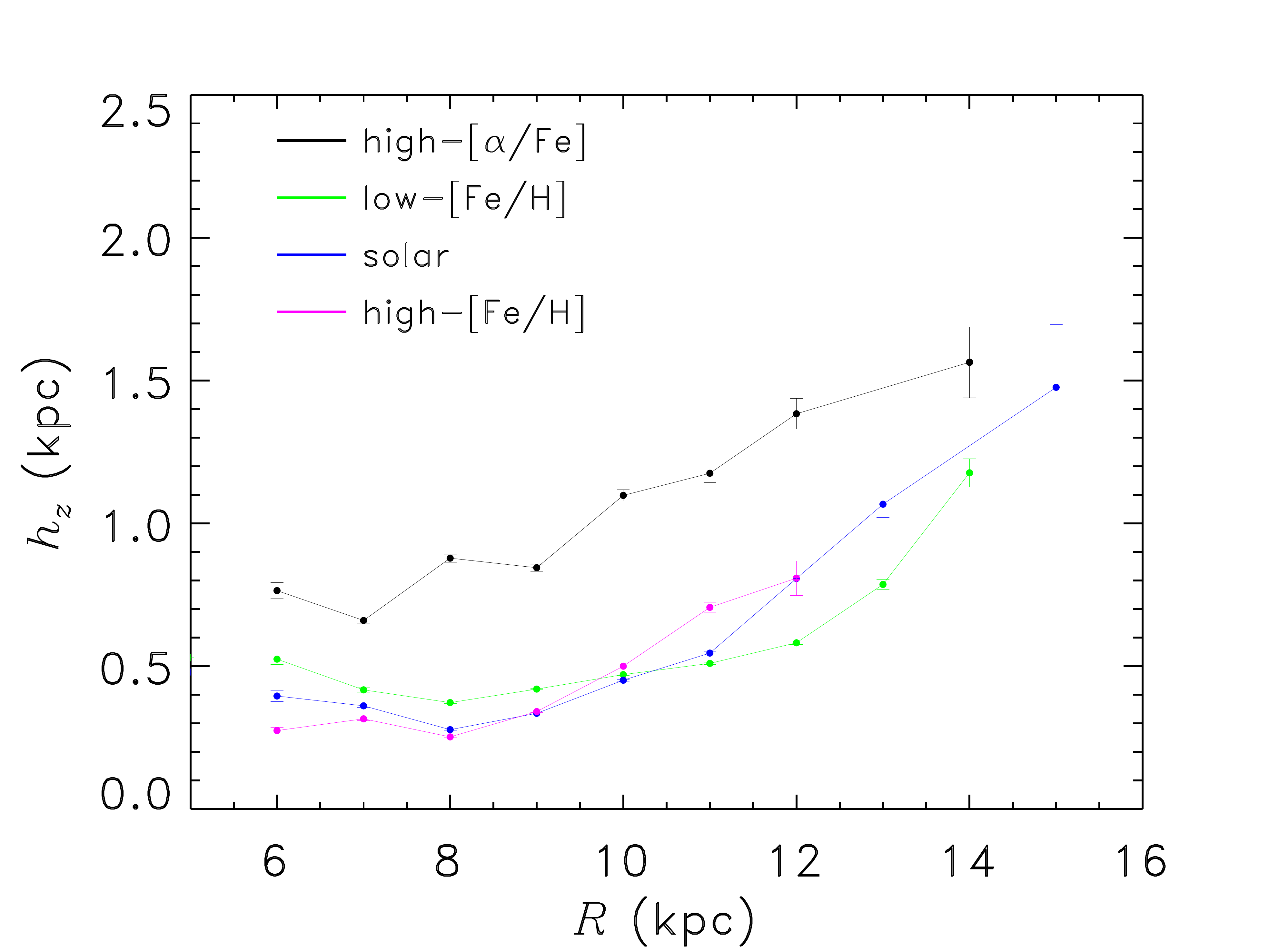}
  \caption{Vertical profiles of the individual MAP. The error bars indicate the 1-sigma uncertainty. This figure displays the radial dependence of the scale height of the individual MAPs as labeled at the top-left corner. \label{fig:f6}}
  \end{figure}

   The vertical density profile distributions of the stellar number and the fitting results by a single exponential for the individual MAPs in different $\vartriangle$$R_{i}$ bins, as labeled in Figure \ref{fig:f5}, and the corresponding $R_{i}$ from inner to outer is  4, 5, 6, 7, 8, 9, 10, 11, 12, 14.5\,kpc, respectively. The shadow regions indicate the distribution ranges of star number density, and the green lines are the best fits determined by the least square method with a $\chi^{2}$-based weighted-mean algorithm.  According to Equation \ref{3}, the inverse of the fitting line slope is the scale height h$_{Z}$ of the individual $\vartriangle$$R_{i}$ bins in each MAP. Consequently, the vertical profiles of the stellar number density can be straightway from the correlation of h$_{Z}$ with $R$ as shown in Figure \ref{fig:f6}.

   Figures \ref{fig:f5} and \ref{fig:f6} show that the vertical stellar number density profiles can be well fitted by a single-exponential for all the four MAPs, but the trends of the vertical profiles are different between the high- and the low-[$\alpha$/Fe] MAPs. The trends of the scale heights along with $R$ are quite similar for all three low-[$\alpha$/Fe] MAPs, with a shallow decline at $R$ $<$ 8\,kpc and a steep rise at $R$ $>$ 8\,kpc. The scale heights reach their minimum values at the same position with $R$ = 8\,kpc. However, the growth trend of $h_{Z}$ for the high-[$\alpha$/Fe] MAP is undulating and relatively moderate compared with the low-[$\alpha$/Fe] MAPs, and its minimum value is at $R$ = 7\,kpc.

\subsection{The Surface-density Profile }\label{sec:hr}
  \begin{figure}
  \includegraphics[height=0.3\textheight,width=0.45\textwidth,clip=]{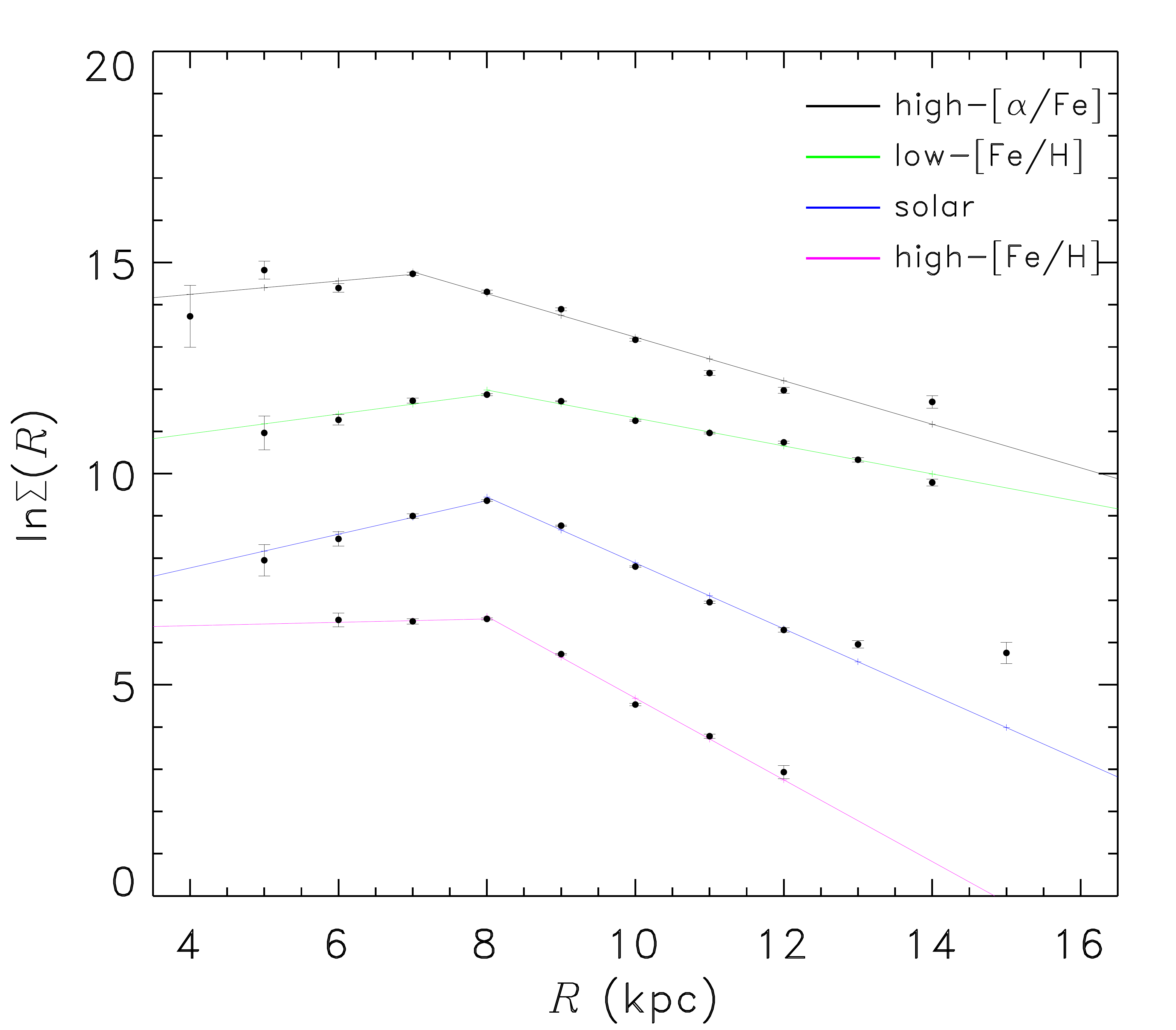}
  \caption{Radial surface profile $\Sigma$($R$) of the four MAPs. An arbitrary offset in the vertical direction has been applied to separate the four profiles. \label{fig:f7}}
  \end{figure}

  B16 have found that the radial surface density profile for their low-[$\alpha$/Fe] MAPs is not a single exponential, but is a much better fit as a broken-exponential profile, rising to a peak radius $R\rm_{peak}$, before falling off.  Although they find a single-exponential fit to the radial profiles of their high-[$\alpha$/Fe] MAPs, they also question the result and think that the high-[$\alpha$/Fe] MAPs maybe have peak radii constrained to be $<$ 5\,kpc, only had not been found due to the coverage limits of their observations. To fit the spatial density profile for sub-samples of our RCs, we follow the methodology of B16. So that we assume a broken exponential model to fit the basic model for the radial surface density profiles $\Sigma$($R$) for all the individual MAPs of our RCs, which is as follows,
    \begin{equation}
    \ln\Sigma(R) \propto \left\{ \begin{array}{ll}
      -h_{R,\mathrm{in}}^{-1} (R-R_0) & R \leq R_{\rm{peak}},\\
      -h_{R,\mathrm{out}}^{-1} (R-R_0) & R > R_{\rm{peak}}.\end{array}\right.
  \label{10}
  \end{equation}
  where $R_{0}$ is the radial distance of the Sun to the Galactic centre on the plane. $h_{R,\rm{in}}$ and $h_{R,\rm{out}}$ are the scale length of the MAPs when $R$ $\leq$ $R_{\rm{peak}}$ and $R$ $>$ $R_{\rm{peak}}$, respectively. $\Sigma$($R$) is the surface density in each radial slice,
  \begin{equation}
  \Sigma(R)=\int_{0}^{+\infty}{2H}dZ=2H_{0}h_{Z}
  \label{11}
  \end{equation}
  where $H$ is the stellar density given by Equation \ref{3}. We also adopt the absolute value $|Z|$ in the above calculations for stars located in southern skies to add the number of stars in each bin and improve the dependability of the fitting functions.

  We calculated the surface stellar number density $\Sigma$($R$) in each radial slice as described by Equation \ref{11}, and the results of radial profile fitting for the four MAPs are displayed in Figure \ref{fig:f7}, where the dots represent the surface number densities in different radial slices as divided in the same way as in Figure \ref{fig:f6}. And the solid lines present the best fit to the dots using the least square method considered the error weight distribution of the stellar number densities in each volume bins.

  Figure \ref{fig:f7} shows that the radial surface number density profiles of all the four MAPs are well fitted by a broken exponential, with a mild upward trend at $R$ $<$ $R\rm_{peak}$ and a sharp downward trend at $R$ $>$ $R\rm_{peak}$. But the peak radius is different between the low- and high-[$\alpha$Fe] MAPs, the three low-[$\alpha$/Fe] MAPs have the same peak radius with $R\rm_{peak}$ = 8\,kpc, while the high-[$\alpha$/Fe] MAP has a smaller peak radius with $R\rm_{peak}$ = 7\,kpc. Moreover, the rising and declining slopes of the radial profiles are also inconsistent with each other of the individual MAPs.

 \begin{table*}
   \newcommand{\mc}[1]{{#1}}
   \centering
   \caption{The comparisons of the disk structure were revealed from APOGEE and LAMOST RC samples. The scale height $h_{Z}$ is the value in solar neighbourhood.   \label{tab:tab}}
   \newcommand{\ch}[1]{\colhead{#1}}
   \newcommand{\dch}[1]{\dcolhead{#1}}
   \newcommand{\ta}{\tablenotemark{a}}
   \newcommand{\tb}{\tablenotemark{b}}
   \begin{tabular}{ l C C C C C C}
   \decimals
   \tableline
     {\rm Sample}&   {\rm MAPs} & {\rm Number} & $h_{R,in}^{-1}$({\rm\,kpc}) & $h_{R,out}^{-1}$({\rm\,kpc}) &$R\rm_{peak}({\rm\,kpc})$& h_{Z}$({\rm\,kpc}) \\
     \tableline
     {\rm B16  } &  {\rm High}-[$\alpha$/{\rm Fe}]  &-&-&  0.43  &-&  0.95     \\
      &   {\rm Low-[Fe/H]}   &-&  0.27   &    0.36   &10.8&   0.37\\
      &   {\rm Solar}   &-&  0.09   &    0.65   &9.4&   0.28\\
      &   {\rm High-[Fe/H]}    &-&   0.28   &   0.81   &6.6&   0.27   \\
   \tableline
     {\rm This work} &  {\rm High}-[$\alpha$/{\rm Fe}]  &11, 137&0.16  &  0.51  &7& 0.88    \\
      &   {\rm Low-[Fe/H]}   &45,403&  0.23  &    0.34   & 8&  0.37\\
      &   {\rm Solar}   &25,484&  0.40   & 0.78   &8& 0.28\\
      &   {\rm High-[Fe/H]}    &8,137&  0.04  &  0.96  & 8&  0.25  \\
   \tableline
   \end{tabular}
   \flushleft
  \end{table*}
\section{Discussion}\label{sec:discussion}
  In this section, we will do some discussion about our results and their implications on the structure and formation of the Galactic stellar disk.

\subsection{ Comparison to B16}\label{sec:compare}
  Since we use the same tracer populations of RCs and the same methodology of mono-abundance populations as used in B16 to investigate the spatial distributions of the different MAPs, it is very significant to compare our results with those of B16.

\subsubsection{The vertical profiles}\label{sec:vertical}
  As illustrated by Figure \ref{fig:f5}, for all the MAPs, the vertical stellar number density distributions in various $\vartriangle$$R_{i}$ bins covered the whole radial range of our RC sample stars are well fitted by a single exponential, which is consistent with the result of B16. Figure \ref{fig:f6} shows that the average scale height h$_{Z}$ at different $\vartriangle$$R_{i}$ bins is a function of $R$, appearing a clear increase of disk thicknesses in the outer disk ($R$ $\geq$ 8\,kpc) for the three low-[$\alpha$/Fe] MAPs. Such an approximately exponential flaring profiles outward for the low-[$\alpha$/Fe] MAPs had been also discovered by B16 in their APOGEE-RC sample (in B16 Figure 13).

  While for the high-[$\alpha$/Fe] MAP, we find a very distinct vertical profile to that  of B16. We find an obvious flaring vertical profile for our high-[$\alpha$/Fe] MAP as shown in Figure \ref{fig:f6}, although the flaring trend is not so sharply as it for the low-[$\alpha$/Fe] MAPs. However, B16 found the thickness of their high-[$\alpha$/Fe] MAPs is constant with $R$ and does not display any flaring. How to explain the difference in vertical profiles of the high-[$\alpha$/Fe] MAPs between this work and B16 is a confusing problem. Because both of our work and B16 use RCs as sample stars, and the same methodology of the mono-abundance population to investigate the stellar populations and the Galactic disk structure. Moreover, both works show the similar metallicity coverage for the high-[$\alpha$/Fe], i.e. $-$1.0 $\leq$ [Fe/H] $\leq$ $-$0.2 in our work and  $-$0.8 $\leq$ [Fe/H] $\leq$ $-$0.2 in B16, and present a very close scaled height as 0.88\,kpc and 0.95\,kpc respectively (see Table \ref{tab:tab}). Therefore, we think that the different vertical profiles most possible result from the differences in the number of RCs and their spatial distribution range of the high-[$\alpha$/Fe] MAP, as well as the values of [$\alpha$/Fe]. First, the number of stars is 11,137 in our high-[$\alpha$/Fe] MAP, which is at least ten times as that in B16 (roughly estimated from the first panel of Figure 6 in B16). Second, the spatial distribution volume of our high-[$\alpha$/Fe] MAP is also larger and more continuous than that of B16 as shown in $R-Z$ plane by Figure \ref{fig:f4} in this work and Figure 6 in B16. Third, the mean value of [$\alpha$/Fe] in our high-[$\alpha$/Fe] MAP is also higher slightly than that of B16 (see Figure 3 in this work and Figure 5 in B16). In fact, B16 had pointed their measurements of the vertical profile for their MAPs are somewhat noisy, because of a lack of data at intermediate and high latitudes, and consequently, it is difficult to obtain truly unbias vertical profiles for the high-[$\alpha$/Fe] MAPs. Moreover, as B16 stated, the high-[$\alpha$/Fe] or old populations have a constant thickness while the low-[$\alpha$/Fe] or younger populations flare significantly has not been seen in any simulation. Most model simulations suggested that the high-[$\alpha$/Fe] population or the thick disk is flaring in the outer disk (e.g., \citealt{Minchev (2014)}; \citealt{Lopez-Corredoira Molgo(2014), Minchev et al.(2015), Minchev et al.(2017)}). For example, \citet{Minchev et al.(2015)} showed that assuming galactic disks formed inside-out, the thick disk is always flare due to the environmental effects and secular evolution, based on two distinct model simulations \citep{Minchev (2013)}. \citet{Minchev et al.(2017)} studied the relationship and equivalence between mono-abundance populations and mono-age populations of the same MAPs of B16 using a chemodynamical model, they found the flaring representative of a mono-age population is only for MAPs at [$\alpha$/Fe] $\geq$ 0.25\,dex (in their Figure 3). They suggested the reason for no flaring in the high-[$\alpha$/Fe] MAPs of B16 considered were centered on [$\alpha$/Fe] = 0.2\,dex (see B16 Figure 5). While our sample stars in high-[$\alpha$/Fe] MAP are mostly at [$\alpha$/Fe] $\geq$ 0.25\,dex (in Figure \ref{fig:f4}) and show a flaring vertical profile as the model predicted by model simulation (e.g., \citealt{Minchev et al.(2017)}).

  In Table \ref{tab:tab} we compare the measurements of the scale lengths and the sale heights in the individual MAPs between our work and B16, including the sample, the inverse scale lengths $h_{R,in}^{-1}$ and $h_{R,out}^{-1}$, the radial profile peak radius $R\rm_{peak}$, and the scale height h$_{Z}$ in the Solar neighbourhood. As Table \ref{tab:tab} listed, the scale heights of the individual MAPs derived from our LAMOST-RC sample are well consistent with those from the B16 APOGEE-RC sample.

\subsubsection{The radial profiles}
  Figure 7 shows that for the three low-[$\alpha$/Fe] MAPs, the radial surface density profiles $\Sigma$($R$) are well fitted by a broken exponential model as suggested by B16 (in their Figure 9), generally with a slow rise at $R$ $<$ $R\rm_{peak}$ and a rapid decline at $R$ $>$ $R\rm_{peak}$. The overall outline of the radial profiles for low-[$\alpha$/Fe] MAPs is very similar to that of B16, but the peak locations of radial surface density for the individual MAPs are different between this work and B16. For our LAMOST-RC sample, the three low-[$\alpha$/Fe] MAPs show a uniform peak radius of $R\rm_{peak}$ = 8\,kpc, while for the APOGEE-RC sample of B16, $R\rm_{peak}$ is variety for their three low-[$\alpha$/Fe] MAPs and increasingly larger for lower [Fe/H] as listed in Table \ref{tab:tab} (also see Figure 9 in B16). B16 explained that such a metallicity dependence of $R\rm_{peak}$ is consistent with the negative radial metallicity gradient of the Galactic thin disk. To match the surface density profiles
  of the APOGEE low-[$\alpha$/Fe] MAPs, \citet{Minchev et al.(2017)} picked their model MAPs with the same central [$\alpha$/Fe] and [Fe/H] values as the low-[$\alpha$/Fe] MAPs defined by B16, based on the analysis for the mono-age populations of the same sample as B16, they suggested that the peak shift of radial profiles for the low-[$\alpha$/Fe] MAPs is due to a bias toward younger ages in the low-[$\alpha$/Fe] MAPs of B16. Because they found that only for stars with age $<$ 4\,Gyr, the low-[$\alpha$/Fe] MAPs should show a shift of $R\rm_{peak}$ to lower radii with increasing of [Fe/H] (see Figure 5 in \citealt{Minchev et al.(2017)}), and they predicted that, for MAPs consist of stars with all ages, the shape of $\Sigma$($R$) will change to mostly inwards of the density peak, while eventually losing the peak for the oldest age groups (see Figure 6 in \citealt{Minchev et al.(2017)}).

\begin{figure}
  \includegraphics[height=0.3\textheight,width=0.45\textwidth,clip=]{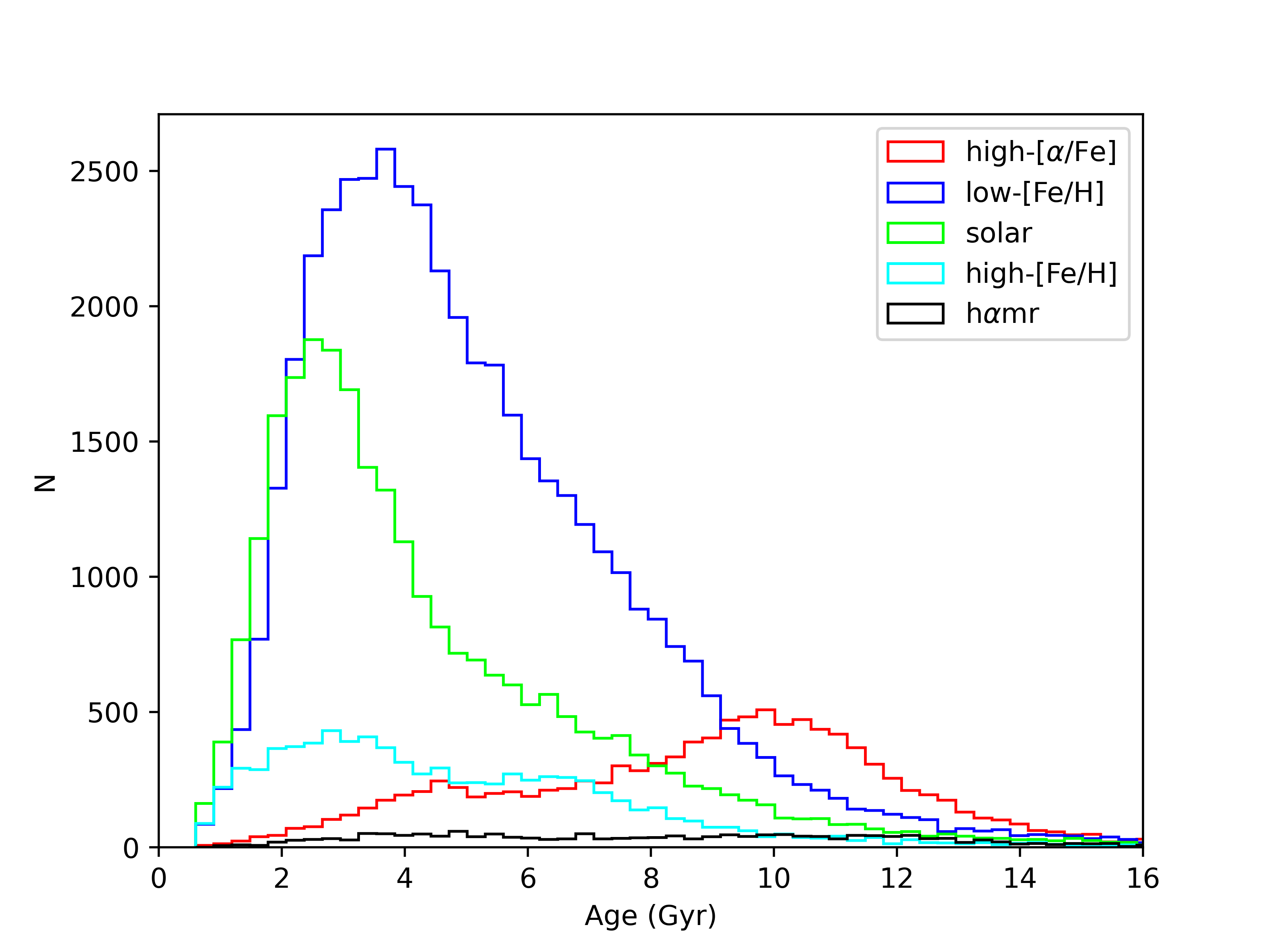}
  \caption{The age distribution of the five MAPs. This distribution shows the age peak of the high-[$\alpha$/Fe] MAP is significantly older than the low-[$\alpha$/Fe] MAPs, see details in the text. \label{fig:f8}}
  \end{figure}

  To investigate the age distributions in the individual MAPs of our RC sample, we plot the age histogram for the individual MAPs in Figure \ref{fig:f8}, including the h$\alpha$mr which will be discussed in subsection \ref{sec:five map}, and the ages of the RCs are from the LAMOST-RC cataloge \citep{Huang et al.(2020)}. Figure \ref{fig:f8} shows that each MAP contains stars with various age. The high-[$\alpha$/Fe] MAP shows two age peaks, the primary age peak is at 10\,Gyr, and the second age peak is at $\sim$ 4.5\,Gyr. The age peak for the low-[Fe/H] is at 4\,Gyr, the solar and high-[Fe/H] MAPs show the same age peak at 2.5\,Gyr. This age peak trend of the different MAPs (except for the h$\alpha$mr MAP)is consistent with the expectation of the typical Age-Metallicity relation. While the h$\alpha$mr MAP does not show any age peak, instead of a flat trend. Figure \ref{fig:f8} also shows that although the age peak for the low-[$\alpha$/Fe] MAPs is concentrated at 2 $-$ 4\,Gyr, each MAP has a significant fraction of stars older than 4\,Gyr especially for the solar MAP and the low-[Fe/H] MAP, which can partly explain the lower $R\rm_{peak}$ for our low-[Fe/H] and solar MAPs than that of B16. So that our results of radial profiles for the low-[$\alpha$/Fe] MAPs are consistent with the model predictions of \citet{Minchev et al.(2017)}. We think that the same $R\rm_{peak}$ for our three low-[$\alpha$/Fe] MAPs implies they are essentially a uniform stellar population, e.g. the thin disk component of the Galaxy, and their $R\rm_{peak}$ is just at the solar radius clearly demonstrates that the Sun is a typical location for the Galactic thin disk.

  For the high-[$\alpha$/Fe] MAP in our LAMOST-RC sample, the radial surface density profile $\Sigma$($R$) also shows a broken exponential similar to the low-[$\alpha$/Fe] MAPs as displayed by Figure \ref{fig:f7}, but with a smaller peak radius of $R\rm_{peak}$ = 7\,kpc. This is in contrast to the APOGEE RC sample of B16, where the high-[$\alpha$/Fe] MAP does not display break or peak in their surface density profiles but is consistent with a single exponential. B16 further suggested their high-[$\alpha$/Fe] MAP could have a peak radius at $R\rm_{peak}$ $\leq$ 4\,kpc and therefore be the continuation of the trend of the low-[$\alpha$/Fe] MAPs, but it is constrained to lie outside of their observed volume. Considering the larger number and larger coverage volume of our high-[$\alpha$/Fe] MAP compared to B16, we prefer to the broken exponential radial density profile for the high-[$\alpha$/Fe] MAP might be true, and the different radial peaks between the high- and low-[$\alpha$/Fe] MAPs indicates that they are distinct components of the Galactic disk. However, the number simulations of the Milky Way chemodynamical model present a single exponential profile to the radial surface density for the high-[$\alpha$/Fe] MAPs \citep{Minchev et al.(2017)}, which is consistent with the results of B16 using the APOGEE RC sample. So our results provide new constraints for the Milky Way chemodynamical model to better match the observed disk structure. Although we present distinct radial profiles for the high-[$\alpha$/Fe] MAPs and different $R\rm_{peak}$ for the low-[$\alpha$/Fe] MAPs as those of B16, the broken exponentials with a $R\rm_{peak}$ for the low-[$\alpha$/Fe] MAPs are consistent with each other. And the corresponding scale lengths of the individual MAPs in out disk are consistent with the measurements of B16 as compared by Table \ref{tab:tab}. While for the inner disk only the low-[Fe/H] MAP presents a well consistent $h_{R}$ compared to B16, and the consistency is poor for the solar and the high-[Fe/H] MAPs.

\subsection{Stellar populations and disk structure}\label{sec:subsec5.2}
\begin{figure}
  \includegraphics[height=0.3\textheight,width=0.45\textwidth,clip=]{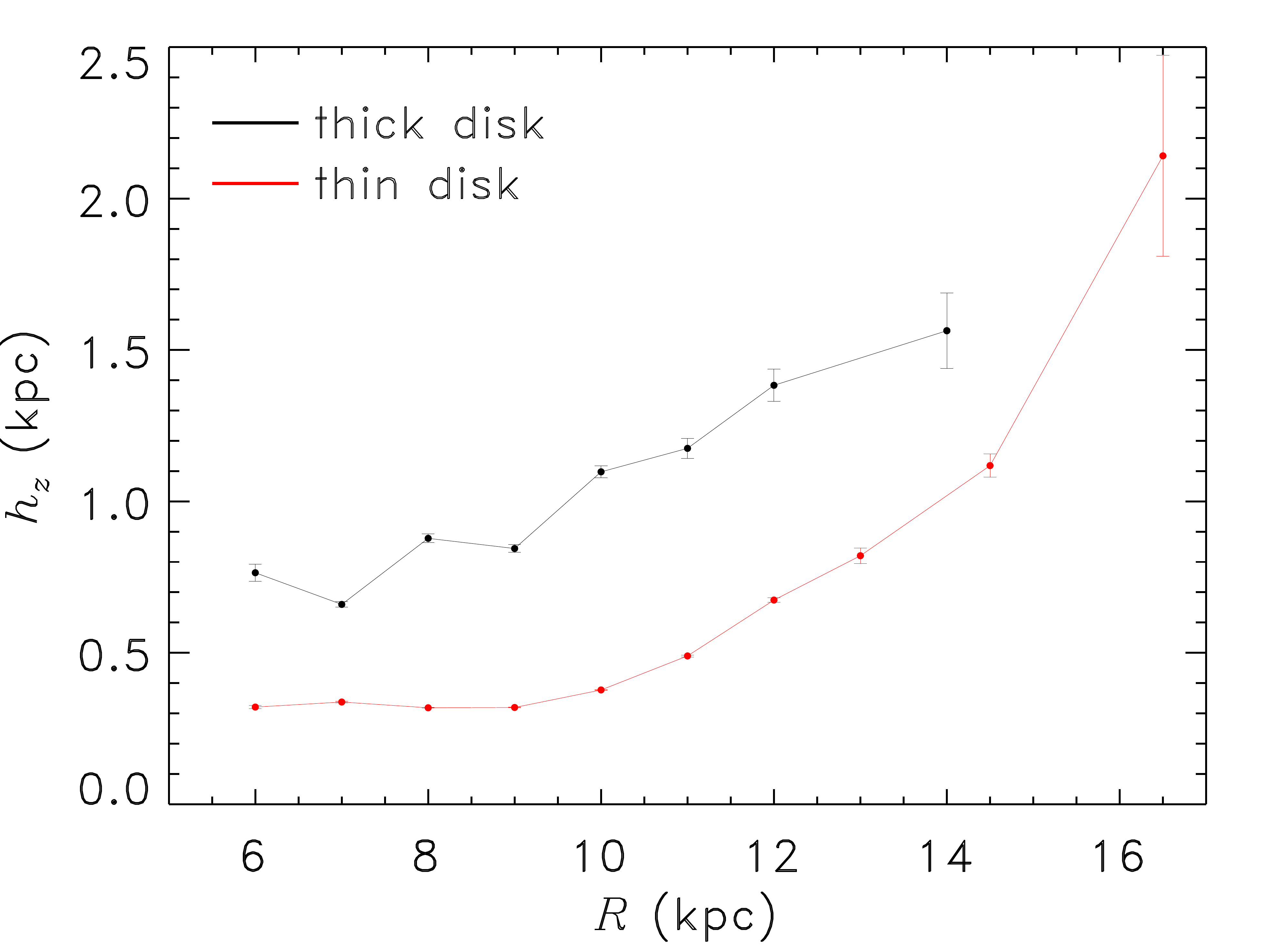}
  \caption{Vertical profiles of the thin- and thick- disks. The error bars indicate the 1-sigma uncertainty. \label{fig:f9}}
  \end{figure}

\begin{figure}
  \includegraphics[height=0.3\textheight,width=0.45\textwidth,clip=]{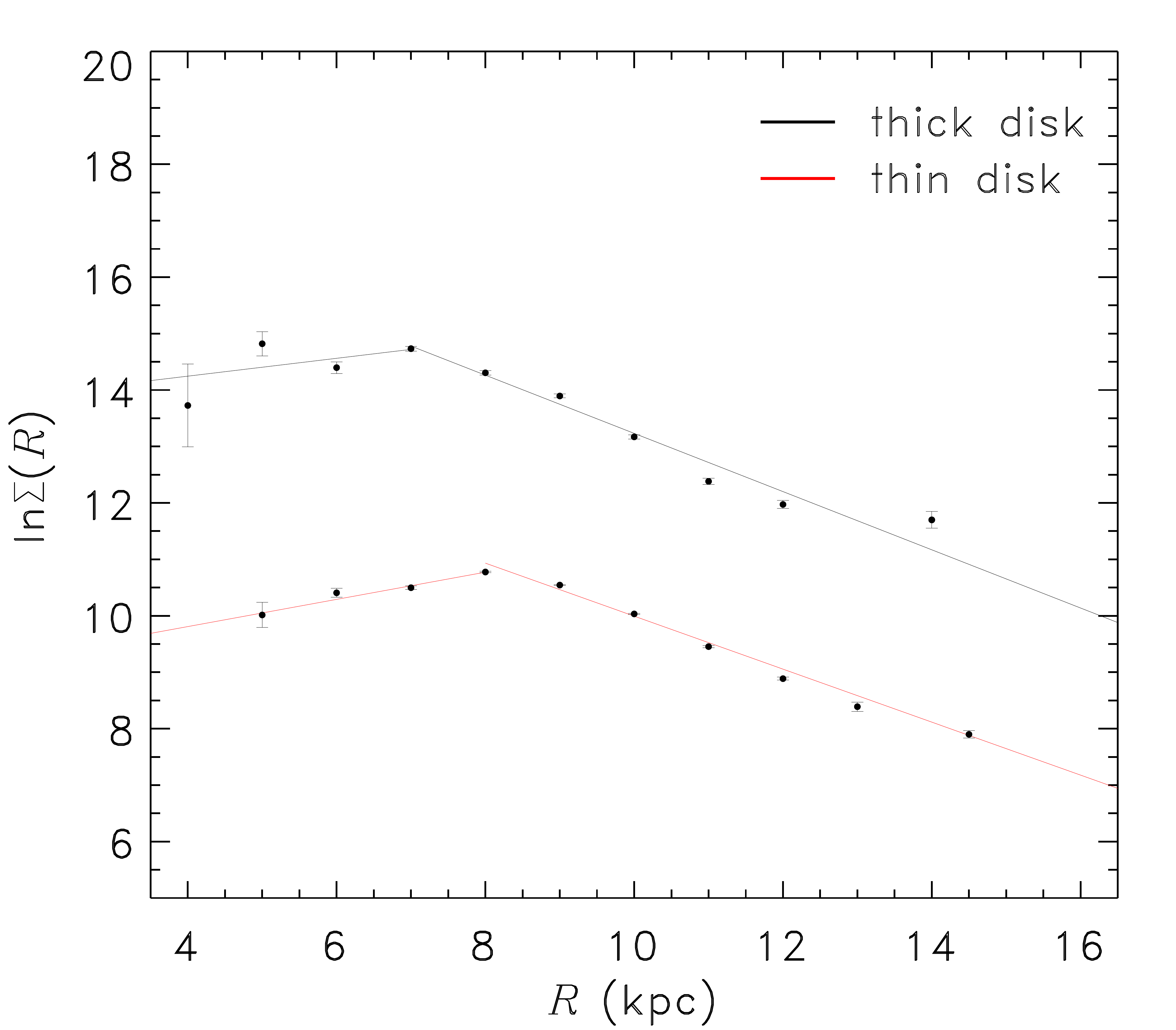}
  \caption{The radial profiles in the thick disk (high-[$\alpha$/Fe] MAP) and the thin disk (low high-[$\alpha$/Fe] MAPs). \label{fig:f10}}
  \end{figure}

  The spatial distributions of the individual MAPs as displayed in Figures \ref{fig:f6} and \ref{fig:f7} have shown that the two separated sequences with high-[$\alpha$/Fe] and low-[$\alpha$/Fe] MAPs are clearly distinct, which means they are two distinct stellar populations. The high-[$\alpha$/Fe] MAP has a larger scale height and a smaller $R\rm_{peak}$ than the low-[$\alpha$/Fe] MAPs. In the meanwhile, the three low-[$\alpha$/Fe] MAPs show quite consistent vertical profiles and the same $R\rm_{peak}$ of the radial profiles. It may thus be concluded that the three low-[$\alpha$/Fe] MAPs actually belong to the same stellar population named as the thin disk of the Galaxy, while the high-[$\alpha$/Fe] MAP corresponds to the thick disk population. To directly compare the spatial distributions of the thin and thick disk populations, we combine the three low-[$\alpha$/Fe] MAPs as the thin disk population and replot the vertical and radial profiles of the two stellar populations in Figures \ref{fig:f9} and \ref{fig:f10} respectively.

  As expected we find that the vertical profile of the thin disk consists of a single exponential, but with a scale height that is flaring outward with an approximately exponential profile from Figure \ref{fig:f9}, which is commonly seen in previous observations (e.g., \citealt{Lopez-Corredoira et al.(2002)}; \citealt{Momany et al.(2006)}; B16; \citealt{Li et al.(2019)}) and simulations of the outer disk (e.g., \citealt{Lopez-Corredoira Molgo(2014)}; \citealt{Kawata et al.(2017)}; \citealt{Agertz et al.(2020)}). The exponentially flaring vertical profile presents a scale height of 0.31\,kpc at solar Galactocentric distance for the thin disk, which is a typical scale height of the thin disk at solar neighbourhood as suggested in literature (e.g., \citealt{Gilmore  Reid(1983)}; \citealt{Juric et al.(2008)}). Although there are some small fluctuations in the vertical profile of the thick disk ( high-[$\alpha$/Fe] MAP ), the increasing trend of thickness with larger Galactocentric distance is significant, and the scale height of 0.88\,kpc in the solar neighbourhood is also in agreement with the typical scale height of thick disk as suggested in literature (e.g., \citealt{Gilmore  Reid(1983)}; \citealt{Juric et al.(2008)}; B16). Another interesting result we find from Figure \ref{fig:f9} is that although the thick disk is always thicker than the thin disk in the whole radial coverage of the high-[$\alpha$/Fe] MAP, the thin disk would be ``thicker" than the thick disk when $R$ $\geq$ 16 kpc if we extrapolate the flaring vertical profile of the  high-[$\alpha$/Fe] MAP to far distance. This result is in good agreement with the simulations of the thin+thick flaring disk model by \citet[see their Figure 5]{Lopez-Corredoira Molgo(2014)}.

  Figure \ref{fig:f10} shows that the radial profiles both for the thin- and thick-disks are well fitted by a broken exponential, with the $R\rm_{peak}$ of 8\,kpc and 7\,kpc for the thin- and thick-disks respectively. From Figure \ref{fig:f10} we calculated an overall scale length of 2.1\,kpc for the thin disk beyond the solar radius, which is larger than the scale length of 1.9\,kpc of the thick disk. Usually, the scale length of the thin disk is larger than 3.0\,kpc and the scale length of the thick disk is a bit larger than 2.0\,kpc \citep{Li2018}. Although our values of the scale lengths are different from those in previous works, the relatively shorter scale length of the thick disk than the thin disk is in agreement with the results in the literatures. For example, \citet{Juric et al.(2008)} present scale lengths of 2.6\,kpc and 3.6\,kpc for the thick and thin disks, respectively. \citet{Bensby et al.(2011)} also found the scale length of the thick disk is much shorter than that of the thin disk, showed as 2.0\,kpc and 3.8\,kpc, respectively. \citet{Li2018} presented the scale length of the thick disk and thin disk obtained by the chemical approach is 2.27\,kpc and the 3.09\,kpc respectively. But the scale length of the thick disk is approximately equal to that of the thin disk via a kinematical approach, which is 2.75\,kpc and 2.76\,kpc for thick and thin disks, respectively. The relative shorter scale length and smaller $R\rm_{peak}$ of the thick disk compared to the thin disk mean that the thick disk is more concentrated in inner disk and the thin disk is more extended in the outer disk, and such a result is a common observational fact (e.g., \citealt{Mikolaitis et al.(2014), Hayden et al.(2015)}; B16).

  In the Milky Way, although flaring in the thin disk is highly favoured, the existence of thick disk flaring in the outskirts is still a matter of debate (e.g., \citealt{Robin et al.(2014), Lopez-Corredoira Molgo(2014)}; B16;  \citealt{Li et al.(2019)}). In any case, there is increasing evidence that thick disk might be more diverse than previously thought in terms of global shape and flaring \citep{Garcia de la Cruz}. For example, using the photometric data from the SDSS-DR8, \citet{Robin et al.(2014)} found that a significant flare appears in the oldest thick disk, but is not found in the younger thick disk. Similarly, the age structures created by the mono-age populations (groups of coeval stars) in thick disk are also showing more diversity for the flaring. \citet{Minchev et al.(2015)} showed that in galactic disks formed inside-out, mono-age populations are well fitted by single exponentials and always flare. In contrast, when the total stellar density is considered, a sum of two exponentials is required for a good fit, resulting in the thin and thick disks, which do not flare (see also in \citealt{Minchev et al.(2017)}). We note that although the mono-age and mono-abundance populations are quite different \citep{Minchev2016}, they present similar flare profiles for the thick disk in the outskirts.

  Our analysis on the LAMOST-RC stars by dissecting the MAPs shows that the chemical bimodality is observed throughout the Galactic disk, and the high- and low-[$\alpha$/Fe] sequences are corresponding to the thick- and thin-disks of the Milky Way, respectively. How to explain the formation mechanism of the stellar thin and thick disks are beyond the scope of this paper, but our results provide some observational constraints to the model of the chemodynamical evolution of the Milky Way disk. Our flared vertical profiles for the thin and thick disks are in good agreement with the predection of thin+thick flaring disk model \citep{Lopez-Corredoira Molgo(2014)}, and are consistent with the number simulations of the chemodynamical evolution in Galactic disks formed in the cosmological context \citep{Minchev (2013), Minchev (2014), Minchev et al.(2015), Minchev et al.(2017)}, as well as the cosmological zoom simulation of VINTERGATAN \citep{Agertz et al.(2020)}. These model simulations suggest that the vertical flaring trends are a natural consequence of inside out, up-side down growth coupled with disk flaring (see also \citealt{Bird et al.(2013), Garcia de la Cruz}), which allows for the low-[$\alpha$/Fe] stars to exist several kpc above the disk's midplane. As analyzed by B16, the exponential flaring profiles for the low-[$\alpha$/Fe] MAPs suggests that radial migration played an important role in the formation and evolution of the thin disk. Radial migration of stars via cold torquing, also known as 'churning', by a bar and spiral waves \citep{Minchev (2013)} then allows for the populations to spatially overlap in the solar neighbourhood. Similar to the flared thin disk, the flaring profile for the high-[$\alpha$/Fe] MAP indicates the radial migration has occurred in the formation of the thick disk as suggested by model simulations (e.g., \citealt{Schonrich  Binney(2009), Minchev et al.(2015), Li2018}). Of course, we cannot rule out the other formation scenarios of the thick disk, such as the accreted gas from satellites \citep{Brook2004}, accreted stars from galaxy mergers \citep{Abadi(2003)}, or from disk-crossing satellites heating up the thin disc \citep{Read(2008)}. On the other hand, the broken exponential radial profiles for the thin- and thick-disks can not be explained by any model of the galactic disks. In fact, nearly all the models we mentioned above present a single exponential profile decreasing with the increasing of $R$ (e.g., \citealt{Minchev et al.(2015), Li2018, Agertz et al.(2020)}). And the smooth downtrend of radial profile in the outer disk ($R$ $>$ $R\rm_{peak}$) as shown in Figure \ref{fig:f10} means that there is no cut-off of the stellar component at $R$ = 14 $-$ 15\,kpc as stated by \citet{Ruphy et al.(1996)}, which is also discovered by B16.


  \begin{figure}
  \includegraphics[height=0.3\textheight,width=0.45\textwidth,clip=]{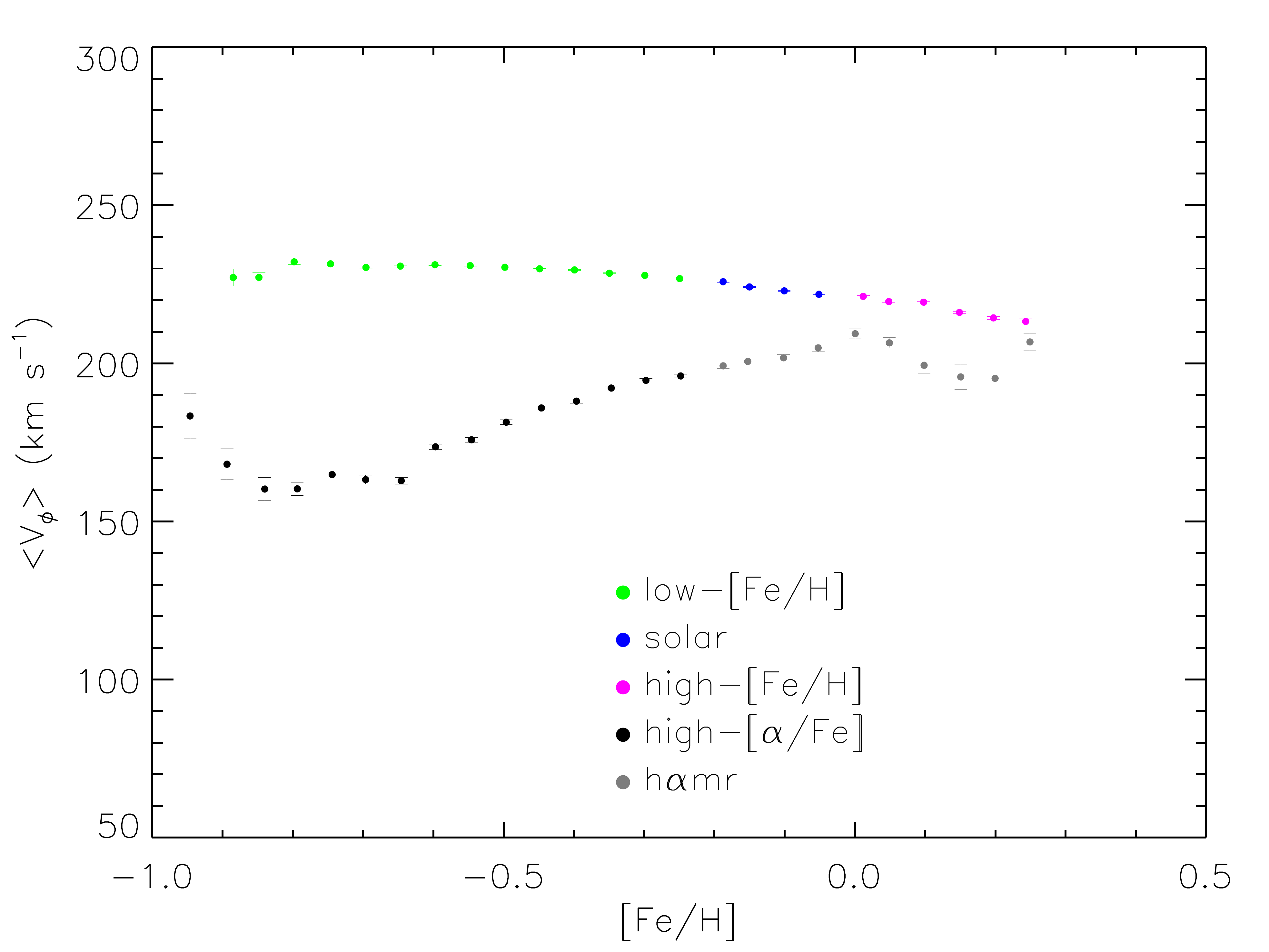}
  \caption{Mean rotational velocity as a function of metallicity for the five MAPs. The dashed line represents the rotation velocity of the local standard
  of rest. \label{fig:f11}}
  \end{figure}

\subsection{Kinematical properties of indivadual MAPs}\label{sec£ºkinematical}

  Applying pure chemical approach based on the [$\alpha$/Fe] vs. [Fe/H] plot as displayed by Figure \ref{fig:f3}, we separate our RCs sample into the thin disk (consisting of the low-[Fe/H], solar, and high-[Fe/H] MAPs with low-[$\alpha$/Fe]), thick disk (the high-[$\alpha$/Fe] MAP), and h$\alpha$mr MAP. Some observational studies have found the different correlations of the mean rotational velocities $V_{\phi}$ with metallicity [Fe/H] between the thin and thick disks \citep{Adibekyan et al.(2013), Wojno et al.(2016), Jia et al.(2018)}. \citet{Adibekyan et al.(2013)} presented a negative gradient of the mean $V_{\phi}$ with [Fe/H] for the thin disk, while it is a positive gradient for the thick disk and the high-$\alpha$ metal-rich stars (see their Figure 4). \citet{Wojno et al.(2016)} also found similar correlations of the mean $V_{\phi}$ with [Fe/H] for their thin and thick disk stars as presented by \citet{Adibekyan et al.(2013)}. To investigate the kinematical properties of the individual MAPs, we plot the trends of the mean $V_{\phi}$ with the metallicity [Fe/H] for each of the five MAPs in Figure \ref{fig:f11}, where the values of the mean $V_{\phi}$ are from the add-value catalogue 2 of LSS-GAC \citep{Xiang et al.(2017b)}, and the data dots with error bars are computed by binning the data into 0.05\,dex wide [Fe/H] bins.

  Figure \ref{fig:f11} shows that, for the high-[$\alpha$/Fe] MAP, the mean $V_{\phi}$ is always smaller than the mean rotation velocity of the Sun, with an increasing trend with metallicity increasing from $-$0.8\,dex to $-$0.2\,dex. For low-[$\alpha$/Fe] MAPs, ignoring the small decline of the mean $V_{\phi}$ when [Fe/H] $<$ $-$0.8\,dex, we find a smoothly continuous descending trend of the mean $V_{\phi}$ with the increasing of [Fe/H] from the low-[Fe/H] MAP to the solar and then to the high-[Fe/H] MAP. This unitary trend of the mean $V_{\phi}$ with [Fe/H] for the three low-[$\alpha$/Fe] MAPs once again indicates that they are actually different components of the thin disk population.

  Our results are in good agreement with those presented by previous works(e.g., \citealt{Adibekyan et al.(2013), Wojno et al.(2016), Jia et al.(2018)}), which means that the low- and high-[$\alpha$/Fe] MAPs separated in chemical abundance distributions are also distinctly separated in kinematical properties. \citet{Adibekyan et al.(2013)} suggested that such a negative correlation between rotational velocities and metallicity for the thin disk stars is a consequence of the radial migration (eg., \citealt{Schonrich  Binney(2009), Loebman et al.(2011)}), which can be explained by inward radial migration of metal-poor and outward migration of metal-rich stars. Interestingly, the trend of the mean $V_{\phi}$ with [Fe/H] for the thick disk (the high-[$\alpha$/Fe] MAPs) as showed in Figure \ref{fig:f11} is amazingly consistent with the result derived from a sample of 147,794 stars of the cross-matching between APOGEE DR14 and Gaia DR2 by \citet[in their Figure 8]{Jia et al.(2018)}, including the complex trends in the metal-poor and metal-rich ends of the whole metallicity range. \citet{Curir et al.(2012)} presented a model of chemical enrichment in the Galaxy by assuming an inverse gradient at high redshifts (z $>$ 3$-$4) explains the positive rotation-metallicity correlation of the thick disk population. They showed that by using the inside-out formation and chemical evolution model of the Galactic disk suggested by \citet{Matteucci  Francois(1989)} and \citet{Chiappini (2001)}, this correlation can be established as a result of radial migration and heating processes of stars from the inner region of the disk.

\subsection{The h$\alpha$mr MAP}\label{sec:five map}

  Many observations have found that there is a kind of disk stars with high-[$\alpha$/Fe] values ($>$ 0.1\,dex) at solar and super-solar metallicities (e.g., \citealt{Gazzano et al.(2013), Adibekyan et al.(2013), Bensby et al.(2014), Jonsson et al.(2018)}). These $\alpha$-enhancements and metal-rich disk stars give valuable insight into the history of gas accretion into our Galaxy \citep{Kordopatis}, but the stellar population property of these stars has not yet been determined. B16 also found the existence of stars with high [$\alpha$/Fe] and high [Fe/H] in their APOGEE-RC sample but they did not discuss stars due to their small number. While in our sample, the h$\alpha$mr MAP consists of 1,746 stars, which is enough to be used as an independent MAP to investigate the membership of stellar population.

  As Figure \ref{fig:f11} illustrated, the trend of the mean $V_{\phi}$ with [Fe/H] for the h$\alpha$mr MAP smoothly continues the increasing trend of the high-[$\alpha$/Fe] MAP for stars with $-$0.2\,dex $\leq$ [Fe/H] $\leq$ 0\,dex, while it becomes decreasing when 0 $\leq$ [Fe/H] $\leq$ 0.2\,dex, and then it returns to increasing when [Fe/H] $>$ 0.2\,dex. The complex up-and-down $V_{\phi}$$-$[Fe/H] trend for the h$\alpha$mr MAP is amazingly consistent with the result of the thick disk presented by \citet[in their Figure 8]{Jia et al.(2018)}, which implies the h$\alpha$mr MAP possibly belongs to the thick disk population. Moreover, our h$\alpha$mr MAP has the same metallicity coverage as the h$\alpha$mr population in \citet[in their Figure 1]{Adibekyan et al.(2013)}, i.e. $-$0.2 $\leq$ [Fe/H] $\leq$ 0.3\,dex. They also found the increasing trend of the mean $V_{\phi}$ with [Fe/H] for their thick disk is continued to their h$\alpha$mr population. So we argue that the h$\alpha$mr MAP is truly belongs to the thick disk population in the kinematical properties.

\begin{figure}
  \includegraphics[height=0.3\textheight,width=0.45\textwidth,clip=]{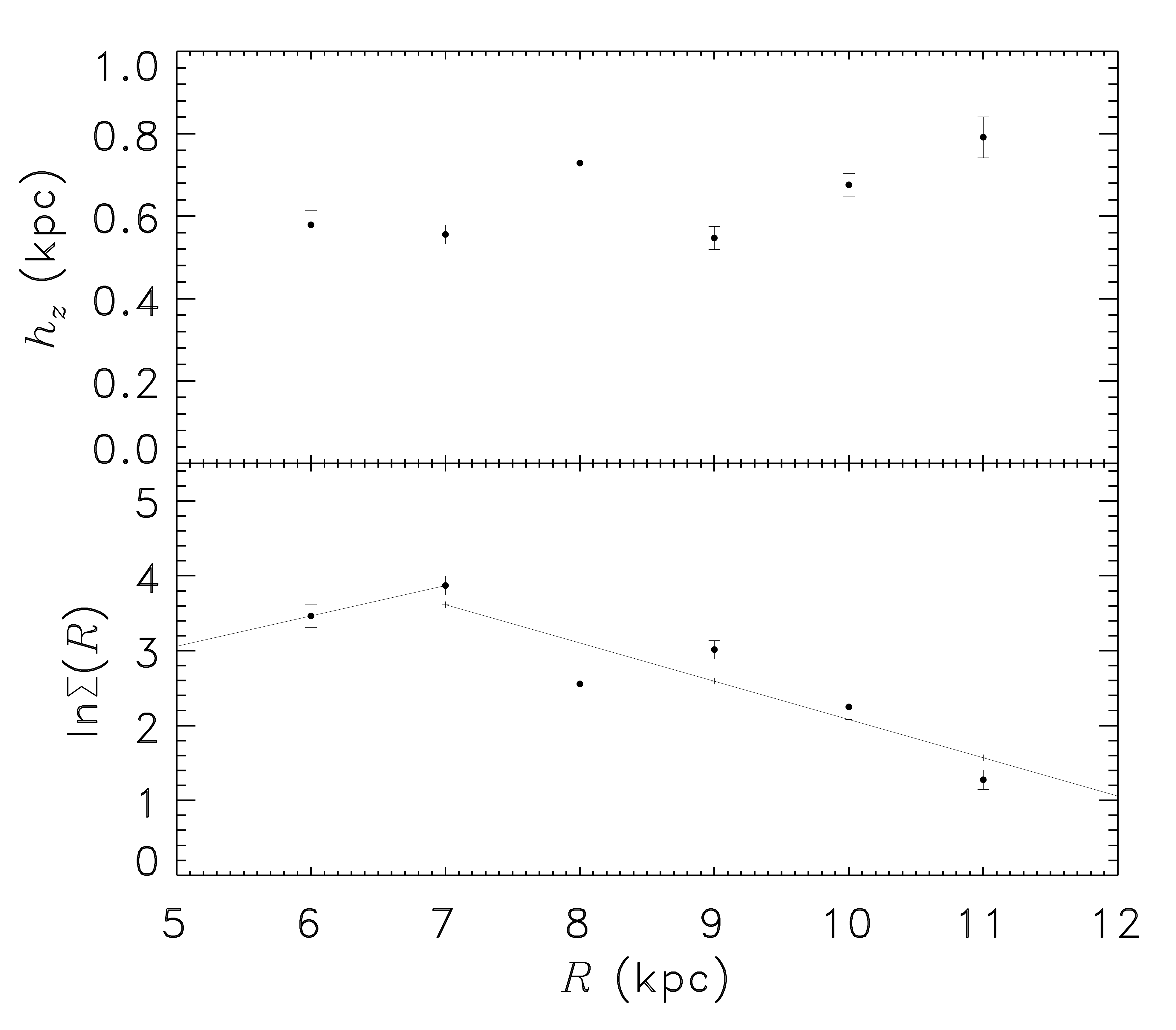}
  \caption{Vertical profile and Radial surface profile of the h$\alpha$mr MAP. The error bars indicate the 1-sigma uncertainty. \label{fig:f12}}
  \end{figure}

  To further determine the stellar population membership of the h$\alpha$mr MAP, we analyzed the vertical and radial stellar number density profiles of this MAP as displayed in Figure \ref{fig:f12}. We note that both the vertical and radial profiles are more similar to those of the high-[$\alpha$/Fe] MAP either in the overall trend or in value of the scale height and the radial density radius peak, although with a relatively small coverage in the radial direction and larger error-bars due to the small number stars in the h$\alpha$mr MAP. So that the spatial distributions of the h$\alpha$mr MAP also prove that it is more likely to belong to the thick disk population. But from the age distributions of the h$\alpha$mr MAP as showed by Figure \ref{fig:f8}, we note there is no age peak but flatly distributed all over the age range of our whole RC sample. Which means that the h$\alpha$mr stars are not mostly made up of the old stars like the thick disk stars.

  Although the h$\alpha$mr stars have equally weighted young and old stars simultaneously, they show the same characteristic as the thick disk stars on the mean $V_{\phi}$$-$[Fe/H] and spatial distributions. Based on the mentioned results, we are inclined to think that the h$\alpha$mr MAP may have originated from the inner Galactic disk and migrated up to the solar neighborhood, although further investigations are needed to clarify their exact nature.

\section{Summary and Conclusions}\label{sec:conclusion}
  Based on a large sample of 96,201 RCs from the LAMOST spectroscopic survey, covering the spatial range of 4\,kpc $\leq$ $R$ $\leq$ 20\,kpc and $-$5\,kpc $\leq$ $Z$$\leq$ 5\,kpc of the Milky Way, we analyzed the distributions of the $\alpha$-elemental abundances in [$\alpha$/Fe]-[Fe/H] and $R$ - $Z$ planes. Then we investigated the distributions of stellar number density both in the Galactic radial and vertical directions for four mono-abundance populations with broad bins in [$\alpha$/Fe] and [Fe/H] and determined the scale heights and scale lengths for the individual MAPs by fitting the vertical and radial stellar number density distributions. We also discussed the correlation of stellar rotation velocity with metallicity for the individual MAPs. Finally, we discussed the population membership of the h$\alpha$mr MAP. Our results confirm that the low-[$\alpha$/Fe] MAPs and the high-[$\alpha$/Fe] MAPs are also separated in spatial distributions and kinematical properties. Our main results are summarized as follows.

\begin{enumerate}[leftmargin=0pt]
\item Our LAMOST RC stars show a clear separation in the [$\alpha$/Fe] versus [Fe/H] plane, the separated low- and high-[$\alpha$/Fe] sequences are also separated in spatial distributions and kinematical properties, corresponding to the typical stellar populations of the thin disk and thick disk, respectively. Moreover, the high-[$\alpha$/Fe] MAP (thick disk) is older than the low-[$\alpha$/Fe] MAPs (thin disk, consisting of the low-[Fe/H], solar, and high-[Fe/H] MAPs) in general.
\item The vertical stellar number density distributions are well fitted by a single exponential both for the thin and thick disks, but show different vertical profiles. The thin disk shows an exponentially flaring profile in the outer disk ($R$ $>$ 9\,kpc) and a flat trend in the inner disk with a scale height of 0.31\,kpc at the solar Galactocentric distance. While the thick disk shows a gradually increasing thickness vertical profile even though with some small fluctuations along with the Galactocentric radius $R$ from inner to outer, and presents a scale height of 0.88\,kpc in the solar neighbourhood. Moreover, the thick disk is always thicker than the thin disk in all over the radial coverage distributed by the high-[$\alpha$/Fe] MAP, but the thin disk would be ``thicker'' than the thick disk if we extrapolate the flaring vertical profile of the thick disk to a far distance at $R$ $\geq$ 16\,kpc.
\item The radial surface number density profiles are well fitted by a broken exponential both for the thin and thick disks, showing a slowly rising (when $R$ $<$ $R\rm_{peak}$) and a rapidly declining (when $R$ $>$ $R\rm_{peak}$) trend. But the thick and thin disks show different values of the peak radius, which is 7\,kpc and 8\,kpc for the thick and thin disks, respectively. The radial profiles show a shorter scale length of 1.9\,kpc for the thick disk than that of 2.1\,kpc for the thin disk in the outer disk (when $R$ $>$ $R\rm_{peak}$), which indicates that the thick disk stars are more centrally concentrated towards the inner Galaxy than the thin disk stars.
\item The thin and thick disks show different relations of the mean rotational velocity $V_{\phi}$ versus metallicity [Fe/H]. The thin disk stars show a negative correlation between the mean $V_{\phi}$ and [Fe/H], with the mean $V_{\phi}$ $>$ 220\,km/s when [Fe/H] $\leq$ 0 and the mean $V_{\phi}$ $\leq$ 220 km/s when [Fe/H] $>$ 0. While the thick disk stars show a positive correlation of the mean $V_{\phi}$ vs. [Fe/H], except for at the most metal-poor end with [Fe/H] $<$ $-$0.85 in which showing a negative correlation. Moreover, the thick disk stars show a mean rotational velocity of smaller than 220\,km/s.
\item There is a special stellar population with high-[$\alpha$/Fe] and high-[Fe/H], i.e. the h$\alpha$mr MAP. It shows similar features to the high-[$\alpha$/Fe] MAP both in the spatial vertical and radial distributions, as well as the correlation of the mean rotational velocity with metallicity. We are inclined to think that the h$\alpha$mr MAP belongs to the thick disk population, and can be regarded as the metal-rich tail of the thick disk.
\item Based on our results as summarized above and model simulations for the Galactic disk, we suggest that the radial migration mechanism is more useful for the formation of the Galactic disk, including the thin and thick disk. As for the h$\alpha$mr MAP, it may have originated from the inner Galactic disk and migrated up to the solar neighborhood.
\item The most difference of our results compared to B16 consists in the spatial distributions of the high-[$\alpha$/Fe] MAP. We present a clear flaring vertical profile but B16 shows a flat vertical profile for the high-[$\alpha$/Fe] MAP. On the other hand, we present a broken exponential for the radial profile for the high-[$\alpha$/Fe] MAP, while B16 shows a single exponential radial profile for them. In addition, we show a common $R\rm_{peak}$ for the radial profiles of the three low-[$\alpha$/Fe] MAPs, instead of the dependence of $R\rm_{peak}$ with [Fe/H] in B16. It is worthwhile to know what these differences mean and further investigations are needed to clarify the exact structure of the Galactic disk.

\end{enumerate}

\acknowledgments
We thank the referee for invaluable suggestion that have greatly improved the manuscript. This work is based on data acquired through the Guoshoujing Telescope. The Guoshoujing Telescope (the Large Sky Area Multi-Object Fiber Spectroscopic Telescope, LAMOST) is a National Major Scientic Project built by the Chinese Academy of Sciences. Funding for the project has been provided by the National Development and Reform Commission. LAMOST is operated and managed by the National Astronomical Observatories, Chinese Academy of Sciences. This work is supported by National Key Basic Research Program of China No. 2014CB845700, National Key R\&D Program of China No. 2019YFA0405502, and National Natural Science Foundation of China U1331120, U1531118 and 11833006. We thank the support of the 2\,m Chinese Space Station Telescope project.

\bibliographystyle{fancyapj}

\end{document}